\documentclass[longauth]{aa}  

\usepackage{graphicx}
\usepackage{mathabx}
\usepackage{txfonts}
\newcommand{\nulup}{$\nu^2$~Lup}
\newcommand{\rearth}{R$_\oplus$}
\newcommand{\mearth}{M$_\oplus$}
\newcommand{\rhoearth}{$\rho_\oplus$}
\newcommand{\rsun}{R$_\odot$}
\newcommand{\msun}{M$_\odot$}
\newcommand{\cheops}{\emph{CHEOPS}}

\begin{document} 

   \title{A full transit of $\nu^2$~Lupi~d and the search for an exomoon \\ in its Hill sphere with  \cheops\thanks{This article uses data from \cheops\ programmes CH\_PR100041 and CH\_PR100031.}}
   \titlerunning{A full transit of $\nu^2$~Lupi~d and the search for an exomoon with \cheops}
   \authorrunning{Ehrenreich et al.}

   \author{
          D.~Ehrenreich\inst{1,2}
          \and
          L.~Delrez\inst{3,4}
          \and
          B.~Akinsanmi\inst{1}
          \and
          T.~G.~Wilson\inst{5}
          \and
          A.~Bonfanti\inst{6}
          \and
          M.~Beck\inst{1}
          \and
          W.~Benz\inst{7,8}
          \and
          S.~Hoyer\inst{9}
          \and
          D.~Queloz\inst{10,11}
          \and
          Y.~Alibert\inst{7}
          \and
          S.~Charnoz\inst{12}
          \and
          A.~Collier Cameron\inst{4}
          \and 
          A.~Deline\inst{1}
          \and
          M.~Hooton\inst{11,7}
          \and
          M.~Lendl\inst{1}
          \and
          G.~Olofsson\inst{13}
          \and
          S.~G.~Sousa\inst{14}
          \and
          V.~Adibekyan\inst{14}
          \and
          R.~Alonso\inst{15}
          \and
          G.~Anglada\inst{16,17}
          \and
          D.~Barrado\inst{18}
          \and
          S.~C.~C.~Barros\inst{14,19}
          \and
          W.~Baumjohann\inst{6}
          \and
          T.~Beck\inst{7}
          \and
          A.~Bekkelien\inst{1}
          \and
          M.~Bergomi\inst{20} 
          \and
          N.~Billot\inst{1}
          \and
          X.~Bonfils\inst{21}
          \and
          A.~Brandeker\inst{13}
          \and
          C.~Broeg\inst{7}
          \and
          T.~B\'arczy\inst{22}
          \and
Z.~K.~Berta-Thompson\inst{50}
          \and
          J.~Cabrera\inst{23}
          \and
          C.~Corral Van Damme\inst{24}
          \and
          S.~Csizmadia\inst{23}
          \and
          M.~B.~Davies\inst{25}
          \and
          M.~Deleuil\inst{9}
          \and
          O.~Demangeon\inst{14,19}
          \and
          B.-O.~Demory\inst{8}
          \and
J.~P.~Doty\inst{48}
          \and
          A.~Erikson\inst{23}
          \and
M.~M.~Fausnaugh\inst{47}
          \and
          H.-G.~Flor\'en\inst{13}
          \and
          A.~Fortier\inst{7}
          \and
          L.~Fossati\inst{6}
          \and
          M.~Fridlund\inst{26,27}
          \and
          D.~Futyan\inst{1}
          \and
          D.~Gandolfi\inst{28}
          \and
          M.~Gillon\inst{3}
          \and
          P.~Guterman\inst{9,29}
          \and
          M.~G\"udel\inst{30}
          \and
          K.~Heng\inst{8,31}
          \and
          K.~G.~Isaak\inst{24}
          \and
          A.~J\"ackel\inst{8}
          \and
J.~M.~Jenkins\inst{49}
          \and
          L.~L.~Kiss\inst{32,33}
          \and
          J.~Laskar\inst{34}
          \and
D.~W.~Latham\inst{46}
          \and
          A.~Lecavelier des Etangs\inst{35}
          \and
A.~M.~Levine\inst{47}
          \and
          C.~Lovis\inst{1}
          \and
          D.~Magrin\inst{36}
          \and
          P.~F.~L.~Maxted\inst{37}
          \and
E.~H.~Morgan\inst{47}
          \and
          V.~Nascimbeni\inst{20}
          \and
H.~P.~Osborn\inst{7,47}
          \and
          R.~Ottensamer\inst{30}
          \and
          I.~Pagano\inst{38}
          \and
          E.~Pall\'e\inst{15}
          \and
          G.~Peter\inst{39}
          \and
          G.~Piotto\inst{20,40}
          \and
          D.~Pollacco\inst{31}
          \and
          R.~Ragazzoni\inst{20,40}
          \and
          N.~Rando\inst{24}
          \and
          H.~Rauer\inst{23,41,42}
          \and
          I.~Ribas\inst{16,17}
          \and
G.~R.~Ricker\inst{47}          
          \and
          S.~Salmon\inst{1}
          \and
          N.~C.~Santos\inst{14,19}
          \and
          G.~Scandariato\inst{38}
          \and
          A.~E.~Simon\inst{7}
          \and
          A.~M.~S.~Smith\inst{23}
          \and
          M.~Steinberger\inst{6} 
          \and
          M.~Steller\inst{6}
          \and
          G.~M.~Szab\'o\inst{43,44}
          \and
          D.~S\'egransan\inst{1}
          \and
A.~Shporer\inst{47}
          \and
          N.~Thomas\inst{7}
          \and
          M.~Tschentscher\inst{23}
          \and
          S.~Udry\inst{1}
          \and
R.~Vanderspek\inst{47}
          \and
          V.~Van Grootel\inst{4}
          \and
          N.~A.~Walton\inst{45}
          }

   \institute{
   Observatoire astronomique de l'Universit\'e de Gen\`eve, chemin Pegasi 51, 1290 Versoix, Switzerland\\ \email{david.ehrenreich@unige.ch}
        \and
   Centre Vie dans l'Univers, Facult\'e des sciences de l'Universit\'e de Gen\`eve, Quai Ernest-Ansermet 30, 1205 Geneva, Switzerland       
        \and
    Astrobiology Research Unit, Universit\'e de Li\`ege, All\'ee du 6 Ao\^ut 19C, B-4000 Li\`ege, Belgium
        \and
    Space sciences, Technologies and Astrophysics Research (STAR) Institute, Universit\'e de Li\`ege, All\'ee du 6 Ao\^ut 19C, 4000 Li\`ege, Belgium
        \and
    Centre for Exoplanet Science, SUPA School of Physics and Astronomy, University of St Andrews, North Haugh, St Andrews KY16 9SS, UK
        \and
    Space Research Institute, Austrian Academy of Sciences, Schmiedlstrasse 6, A-8042 Graz, Austria
        \and
    Physikalisches Institut, University of Bern, Sidlerstrasse 5, 3012 Bern, Switzerland
        \and
    Center for Space and Habitability, University of Bern, Gesellschaftsstrasse 6, 3012 Bern, Switzerland
        \and
    Aix Marseille Univ, CNRS, CNES, LAM, 38 rue Fr\'ed\'eric Joliot-Curie, 13388 Marseille, France
        \and
    ETH Zurich, Department of Physics, Wolfgang-Pauli-Strasse 2, CH-8093 Zurich, Switzerland
        \and
    Cavendish Laboratory, JJ Thomson Avenue, Cambridge CB3 0HE, UK
        \and
    Université de Paris, Institut de physique du globe de Paris, CNRS, F-75005 Paris, France
        \and
    Department of Astronomy, Stockholm University, AlbaNova University Center, 10691 Stockholm, Sweden
        \and
    Instituto de Astrofisica e Ciencias do Espaco, Universidade do Porto, CAUP, Rua das Estrelas, 4150-762 Porto, Portugal
        \and
    Instituto de Astrofisica de Canarias, 38200 La Laguna, Tenerife, Spain ; Departamento de Astrofisica, Universidad de La Laguna, 38206 La Laguna, Tenerife, Spain
        \and
    Institut de Ciencies de l'Espai (ICE, CSIC), Campus UAB, Can Magrans s/n, 08193 Bellaterra, Spain
        \and
    Institut d'Estudis Espacials de Catalunya (IEEC), 08034 Barcelona, Spain
        \and
    Depto. de Astrofisica, Centro de Astrobiologia (CSIC-INTA), ESAC campus, 28692 Villanueva de la Cañada (Madrid), Spain
        \and
    Departamento de Fisica e Astronomia, Faculdade de Ciencias, Universidade do Porto, Rua do Campo Alegre, 4169-007 Porto, Portugal
        \and
    INAF, Osservatorio Astronomico di Padova, Vicolo dell'Osservatorio 5, 35122 Padova, Italy
        \and
    Université Grenoble Alpes, CNRS, IPAG, 38000 Grenoble, France
        \and
    Admatis, 5. Kandó Kálmán Street, 3534 Miskolc, Hungary
        \and
    Institute of Planetary Research, German Aerospace Center (DLR), Rutherfordstra\ss e 2, 12489 Berlin, Germany      \and
    Science and Operations Department - Science Division (SCI-SC), Directorate of Science, European Space Agency (ESA), European Space Research and Technology Centre (ESTEC), Keplerlaan 1, 2201-AZ Noordwijk, The Netherlands
        \and
    Centre for Mathematical Sciences, Lund University, Box 118, 221 00 Lund, Sweden
        \and
    Leiden Observatory, University of Leiden, PO Box 9513, 2300 RA Leiden, The Netherlands
        \and
    Department of Space, Earth and Environment, Chalmers University of Technology, Onsala Space Observatory, 439 92 Onsala, Sweden
        \and
    Dipartimento di Fisica, Universita degli Studi di Torino, via Pietro Giuria 1, I-10125, Torino, Italy
        \and
    Division Technique INSU, CS20330, 83507 La Seyne-sur-Mer cedex, France
        \and
    Department of Astrophysics, University of Vienna, T\"urkenschanzstrasse 17, 1180 Vienna, Austria
        \and
    Department of Physics, University of Warwick, Gibbet Hill Road, Coventry CV4 7AL, United Kingdom
        \and
    Konkoly Observatory, Research Centre for Astronomy and Earth Sciences, 1121 Budapest, Konkoly Thege Miklós út 15-17, Hungary
        \and
    ELTE E\"otv\"os Lor\'and University, Institute of Physics, P\'azm\'any P\'eter s\'et\'any 1/A, 1117 Budapest, Hungary
        \and
    IMCCE, UMR8028 CNRS, Observatoire de Paris, PSL Univ., Sorbonne Univ., 77 av. Denfert-Rochereau, 75014 Paris, France
        \and
    Institut d'astrophysique de Paris, UMR7095 CNRS, Université Pierre \& Marie Curie, 98bis blvd. Arago, 75014 Paris, France
        \and
    INAF, Osservatorio Astronomico di Padova, Vicolo dell'Osservatorio 5, 35122 Padova, Italy
        \and
    Astrophysics Group, Keele University, Staffordshire, ST5 5BG, United Kingdom
        \and
    INAF, Osservatorio Astrofisico di Catania, Via S. Sofia 78, 95123 Catania, Italy
        \and
    Institute of Optical Sensor Systems, German Aerospace Center (DLR), Rutherfordstrasse 2, 12489 Berlin, Germany
        \and
    Dipartimento di Fisica e Astronomia ``Galileo Galilei", Universit\`a degli Studi di Padova, Vicolo dell'Osservatorio 3, 35122 Padova, Italy
        \and
    Zentrum f\"ur Astronomie und Astrophysik, Technische Universit\"at Berlin, Hardenbergstr. 36, D-10623 Berlin, Germany
        \and
    Institut f\"ur Geologische Wissenschaften, Freie Universit\"at Berlin, 12249 Berlin, Germany
        \and
    ELTE E\"otv\"os Lor\'and University, Gothard Astrophysical Observatory, 9700 Szombathely, Szent Imre h. u. 112, Hungary
        \and 
    MTA-ELTE Exoplanet Research Group, 9700 Szombathely, Szent Imre h. u. 112, Hungary
        \and
    Institute of Astronomy, University of Cambridge, Madingley Road, Cambridge, CB3 0HA, United Kingdom
        \and
    Harvard-Smithsonian Center for Astrophysics, 60 Garden Street, Cambridge, MA 02138, USA
        \and
    Department of Physics and Kavli Institute for Astrophysics and Space Research, Massachusetts Institute of Technology, 77 Massachusetts Avenue, Cambridge, MA 02139, USA
        \and
    Noqsi Aerospace Ltd., 15 Blanchard Avenue, Billerica, MA 01821, USA 
        \and
    NASA Ames Research Center, Moffett Field, CA 94035, USA 
        \and
    Department of Astrophysical and Planetary Sciences, University of Colorado, Boulder, CO 80309, USA
     }

   \date{}

  \abstract{
  The planetary system around the naked-eye star $\nu^2$~Lupi (\object{HD~136352}; TOI-2011) is composed of three exoplanets with masses of 4.7, 11.2, and 8.6~Earth masses (\mearth). The \emph{TESS} and \cheops\ missions revealed that all three planets are transiting and have radii straddling the radius gap separating volatile-rich and volatile-poor super-earths. Only a partial transit of planet d had been covered so we re-observed an inferior conjunction of the long-period 8.6~\mearth\ exoplanet \nulup~d with the \cheops\ space telescope. We confirmed its transiting nature by covering its whole 9.1~h transit for the first time. We refined the planet transit ephemeris to $P = 107.1361^{+0.0019}_{-0.0022}$~days and $T_c = 2\,459\,009.7759^{+0.0101}_{-0.0096}$~$\rm BJD_{TDB}$, improving by $\sim40$ times on the previously reported transit timing uncertainty. This refined ephemeris will enable further follow-up of this outstanding long-period transiting planet to search for atmospheric signatures or explore the planet's Hill sphere in search for an exomoon. In fact, the \cheops\ observations also cover the transit of a large fraction of the planet's Hill sphere, which is as large as the Earth's, opening the tantalising possibility of catching transiting exomoons. We conducted a search for exomoon signals in this single-epoch light curve but found no conclusive photometric signature of additional transiting bodies larger than Mars. Yet, only a sustained follow-up of $\nu^2$~Lup~d transits will warrant a comprehensive search for a moon around this outstanding exoplanet.
  }

   \keywords{Planets and satellites: individual: $\nu^2$~Lupi --- techniques: photometric}

   \maketitle
%

\section{Introduction}

   Exoplanets with masses intermediate between that of terrestrial planets and icy giants are mysterious objects. Some, dubbed `rocky super-earths', have bulk densities compatible with the composition of terrestrial planets \citep{mayor2011,dressing2015} and could thus be scaled-up versions of Earth or Mercury \citep{valencia2006,elkinstanton2008,dorn2019}. Other, larger, objects have lower densities that are compatible with the presence of a volatile (ice or steam) or gas (hydrogen and helium) envelope. Depending on authors, these have been called `ocean-planets' , `volatile-rich super-earths', `mini-neptunes' or `sub-neptunes' \citep{kuchner2003,leger2004,santos2004,grasset2009}; the lack of a universally accepted designation illustrates their uncertain nature. These two groups of objects appear to be separated by a radius gap in the exoplanet population \citep{fulton2017,fulton2018}, which indicates different histories \citep{owen2017}: formerly gas- or volatile-rich planets could have lost their envelopes via photoevaporation or core-powered mass-loss to become volatile-poor planets \citep{venturini2020,rogers2021}. Alternatively, the latter could have formed intrinsically devoid of ice or gas.
   
   The origin of rocky super-earths as a product of the evolution of sub-neptunes largely depends on the level of high-energy stellar irradiation received by the planets, which is driving the escape of their gaseous envelope \citep{lammer2003,lecaveier2007}. However, the planet irradiation history is challenging to reconstruct and tricky to compare across different systems. In this respect, multi-planet systems are critically important; they give access to exoplanets with irradiation history that can be compared accurately. In addition, when planets in such system transit their stars, it becomes possible to measure their densities and surface gravities, and probe their atmospheres. This provides inestimable insights into their present-day composition and, by comparing them together, their past evolution \citep{lopez2012}. 
   
   The system of three planets around the naked-eye star $\nu^2$~Lupi \citep[\nulup, also known as \object{HD~136352}, HR~5699 or TOI-2011;][]{udry2019} offers one of the best known opportunities for comparative studies of super-earth evolution. $\nu^2$~Lupi is a close-by (14.7~pc), old ($11.7\pm2$~Gyr) solar-type (G4{\sc v}) star (see Table~\ref{tab:star}). Its three planets provide huge dynamics in terms of stellar irradiation, ranging from $\sim$100 times the insolation of the Earth at planet b to $\sim$5 times at planet d, which could have been spared by intense photoevaporation, 0.425~au away from its star (between Mercury and Venus in the Solar System), and still possess a low gas content of primordial origin \citep{delrez2021}.

   Interest in the \nulup\ system exploded when the \emph{Transiting Exoplanet Survey Satellite (TESS)} detected transits by the two inner planets b and c \citep{kane2020}. Planet b has an orbital period of 11.6~d and a mass of $4.72\pm0.42$~Earth masses (\mearth), while planet c is $11.24^{+0.65}_{-0.63}$~\mearth\ at 27.6~d, which is relatively long for a transiting planet. \citet{kane2020} measured largely different densities of $7.8^{+1.2}_{-1.1}$~g~cm$^{-3}$ for b and $3.50^{+0.41}_{-0.36}$~g~cm$^{-3}$ for c, placing these planets on opposite sides of the radius valley.
   
   \citet{delrez2021} followed up the transits of \nulup\ b and \nulup\ c with the \emph{Characterising Exoplanets Satellite (CHEOPS)}. Their observations (see their Figs.~1 and~S1) serendipitously revealed a third transit-like event, attributed to planet d passing in front of the star altogether with planet c. This was a fortunate finding, because planet d -- with a mass of $8.82^{+0.93}_{-0.92}$~\mearth\ -- has a 107~d period that made its transit unlikely to be seen. However, \citet{delrez2021} showed that the transit-like event is compatible with the radial-velocity-based (hence, relatively loose) ephemeris for planet d, has a depth compatible with a planet of $\sim2.6$~Earth radii (\rearth), and exhibits an in-transit curvature consistent with the known limb-darkening profile of the host star; all these elements are in favour of planet d being transiting in spite of an incomplete observation: \cheops\ stopped collecting data before the egress. As a result, the transit duration and mid-transit time are loosely constrained. This makes any follow-up attempts of this unusually long-period transiting planet risky, whilst the scarce opportunities are costly in observing time (the transit duration is estimated to be longer than 8~h). A robust ephemeris is sorely needed to achieve the follow-up \nulup\ d deserves.
   
\begin{figure*}
    \centering
    \resizebox{\textwidth}{!}{\includegraphics{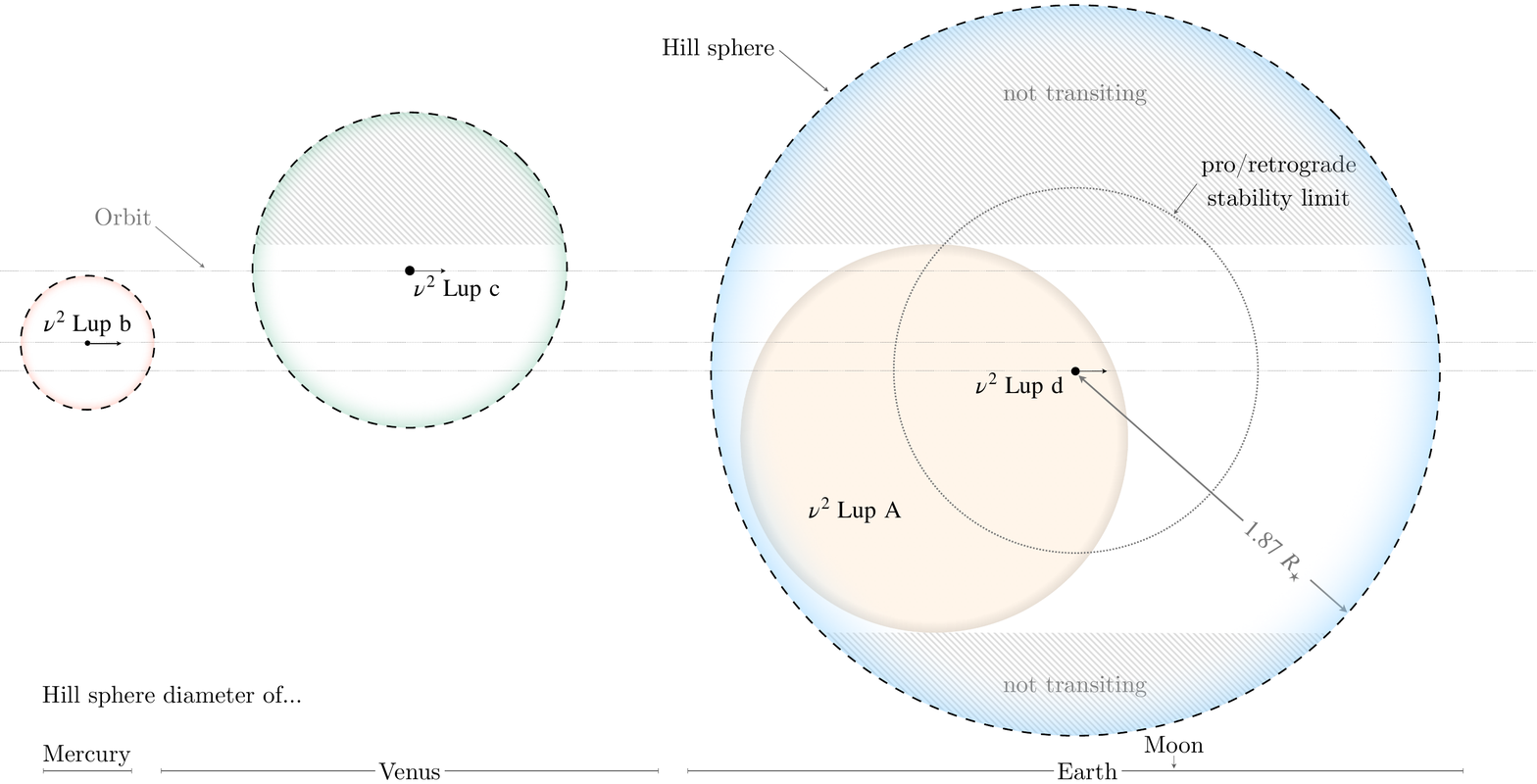}}
    \caption{Comparison of the sizes of the Hill spheres (dashed circles) for the three \nulup\ planets. Planet~d is represented while transiting the star (ivory white disc). The respective orbital positions of the planets are not representative of their positions during the observations; only their transit impact parameters are considered. The Hill sphere radius of \nulup~d (1.87~$R_\star$ or 0.0092~au or 9.1~h) is almost equal to the length of the planet transit chord. Because the Hill sphere of \nulup~d is larger than the star, some parts of it (hatched areas) do not transit: a highly inclined exomoon in these regions, close to maximum apparent separation, could remain undetectable during this single visit. Such a distant moon would most likely be on a stable retrograde orbit, as these non-transiting regions are mostly beyond the stability limit for prograde (circular) orbits, represented at half the Hill sphere radius \citep{domingos2006} by the dotted circle. The diameters of the Hill spheres of Mercury, Venus, and Earth are shown for comparison; the location of the Moon within Earth's Hill sphere is indicated.}
    \label{fig:hillspheres}
\end{figure*}

   Here, we report on a second-epoch \cheops\ observation specifically designed to confirm the transit of \nulup\ d and to obtain precise transit parameters and a reliable ephemeris. The analysis of the observations (Sect.~2) shows a clear detection and confirmation of a complete transit of \nulup\ d, enabling us to derive a precise ephemeris and refine the planet properties (Sect.~3). Furthermore, this 8.6~\mearth\ planet located relatively far from its host star has a Hill sphere similar in size to that of the Earth (Fig.~\ref{fig:hillspheres}), allowing for the presence of satellites that could transit alongside the planet. As our observations cover the full transit of the Hill sphere of \nulup~d, we investigate the presence of moons in Sect.~4.

\section{Observations and data reduction}

\subsection{New data}

\subsubsection{\cheops}
\label{sec:cheopsDataReduction}

\cheops\ \citep{benz2021} initially observed \nulup\ in its early science programme (CH\_PR100041), which was part of the guaranteed time observations (GTO), from April to July 2020. \nulup\ d was first detected in transit on 9 June 2020 \citep{delrez2021}. The first window of opportunity for a confirmation was in late April 2021. We scheduled a new \cheops\ visit from 24 April to 28 April 2021 to cover both the $3\sigma$ window on the transit-timing uncertainty and the transit of the planet Hill sphere, as part of the GTO programme CH\_PR100031. The visit represents 56 \cheops\ orbits (3.85 days). 

\begin{figure*}[!h]
    \centering
    \resizebox{\textwidth}{!}{\includegraphics{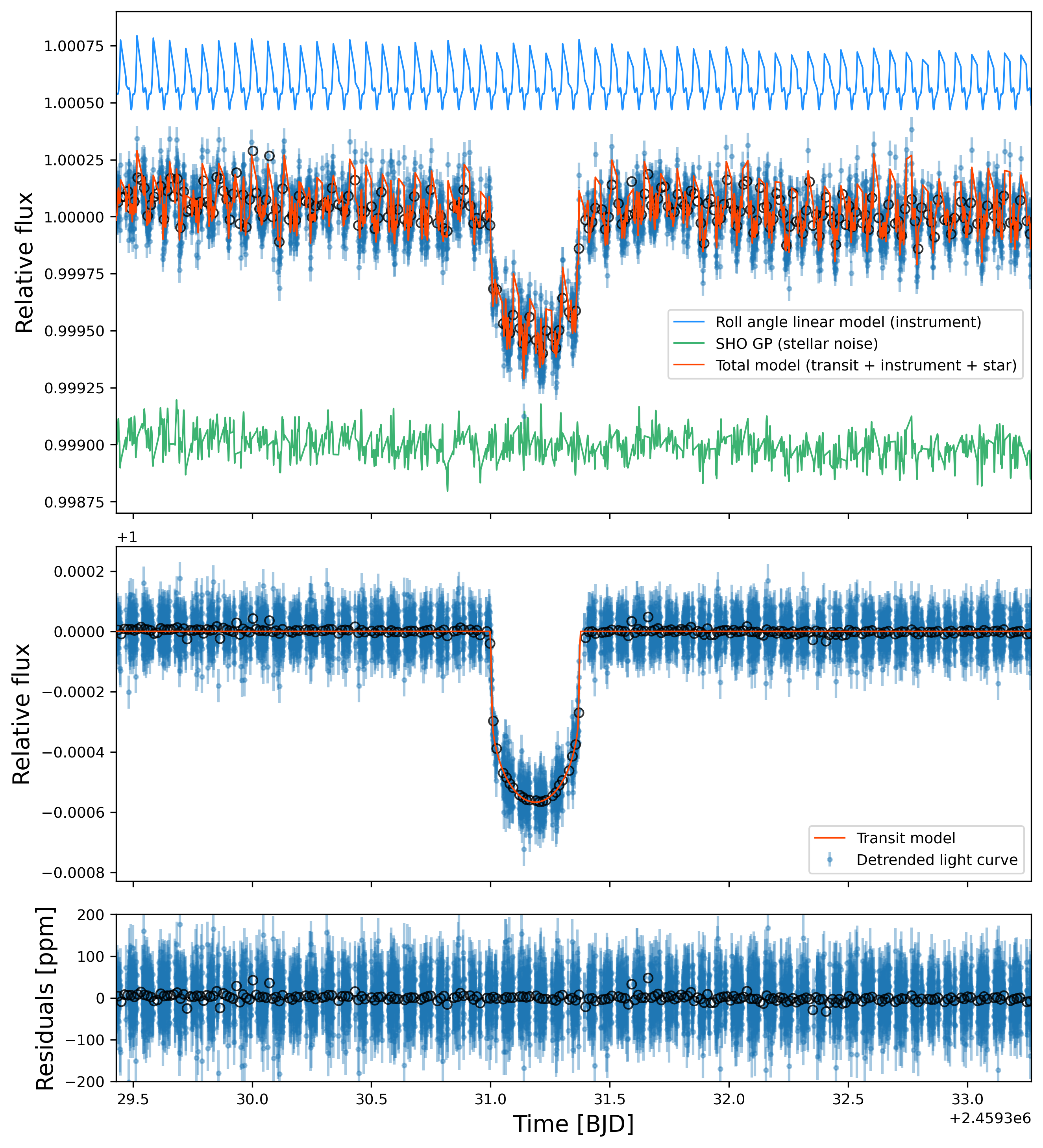}}
    \caption{New \cheops\ transit light curve of \nulup\ d. \emph{Top panel:} The light curve reduced with the \cheops\ automated \texttt{DRP} (blue points with error bars), modelled with a combination of instrumental effects (blue curve), stellar noise (green curve), and a transit. The model including all these effects is shown in orange. \emph{Middle panel:} Light curve corrected for the instrumental and stellar noise models (blue points), together with the best-fit transit model (orange curve). See Fig.~\ref{fig:lc_zoom} for a zoom on the transit itself. \emph{Bottom panel:} Residuals resulting from the difference between the detrended light curve and the transit model. For all panels, the error bars of the data points include the fitted additional jitter term added in quadrature. The open black circles show the light curve binned into 20-minute intervals.}
    \label{fig:lc_d_CHEOPS}
\end{figure*}

A total of 4\,720 images were obtained; each image is a 200$\times$200-pixel subarray resulting from the co-addition, performed on board, of 26 exposures of 1.7~s; one such image is obtained every 44.2~s. Images were processed by the \cheops\ data reduction pipeline \citep[\texttt{DRP v13.1.0};][]{hoyer2020}. The \texttt{DRP} calibrates the raw images (event flagging, bias and gain corrections, linearisation, dark current, and flat-field corrections), corrects for environmental effects (smearing trails, jitter, background, and stray light), and performs a photometric extraction using several circular apertures \citep{hoyer2020}, out of which we selected the default aperture size (25 pixels in radius), following \citet{delrez2021}. The output is the photometric time series reproduced in the upper panel of Fig.~\ref{fig:lc_d_CHEOPS}. Contamination of the photometric aperture by background stars is automatically estimated by the \texttt{DRP} and is typically small (0.027-0.031\%) for these observations of \nulup. The resulting light curve features interruptions due to Earth occultations and passages throughout the South Atlantic Anomaly. The \cheops\ light curve has an RMS of 48~ppm (unbinned cadence of 44.2~s), 28~ppm (binned by 2 min), and 9~ppm (binned by 20 min).

\begin{figure}[!h]
    \centering
    \resizebox{\columnwidth}{!}{\includegraphics{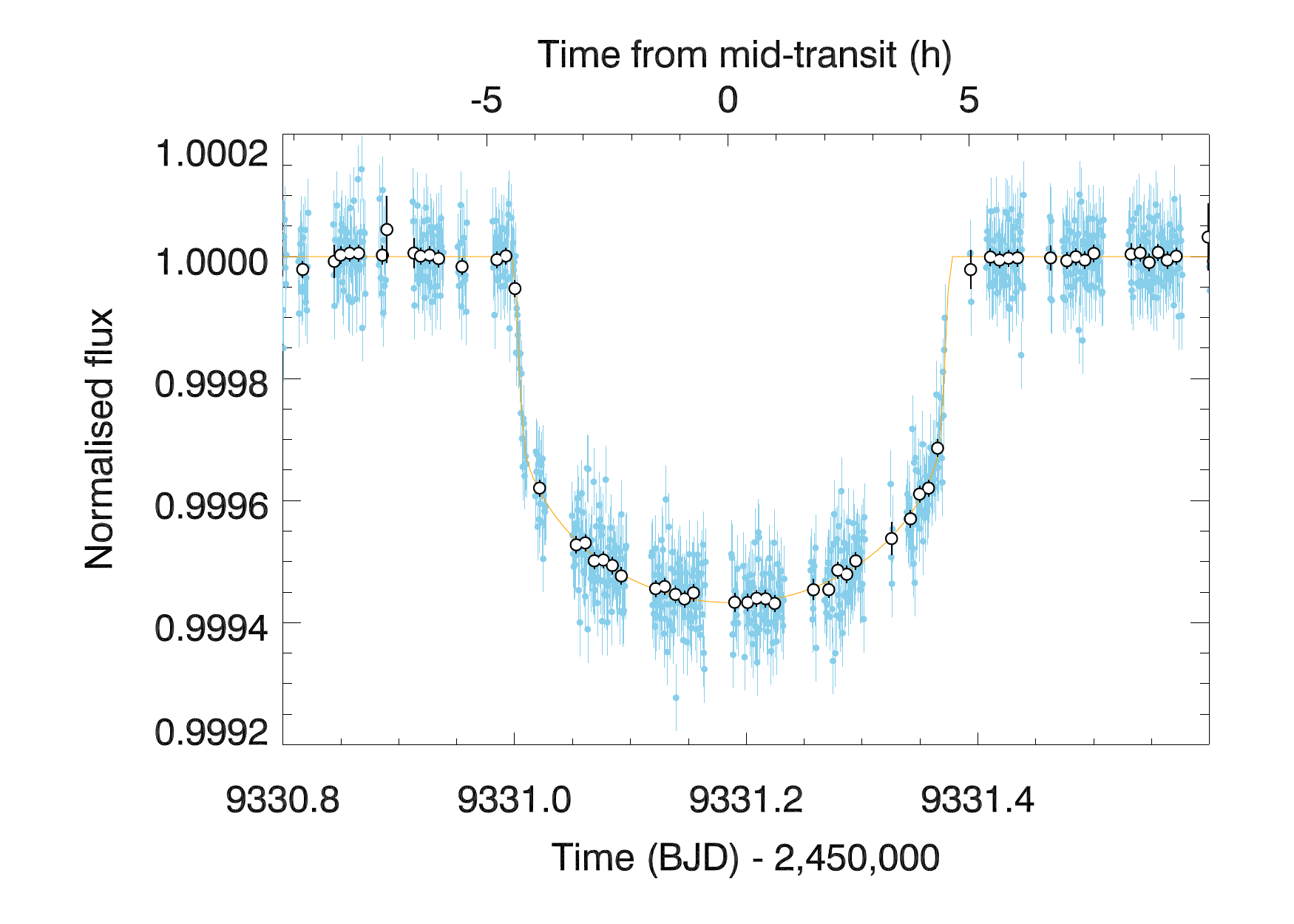}}
    \caption{Zoom on the detrended transit of \nulup~d from the middle panel of Fig.~\ref{fig:lc_d_CHEOPS}. Unbinned exposures are shown as sky blue points with error bars, and white points show a binning by a factor of 15 (or lower depending on how many exposures there are between two light-curve interruptions). The transit model is represented by an orange curve.}
    \label{fig:lc_zoom}
\end{figure}

\subsubsection{\emph{TESS}}
\emph{TESS} first observed \nulup\ during sector~12 of its primary mission, between 21 May 2019 and 18 June 2019. The transits of planets b and c were discovered in these observations \citep{kane2020}, which did not cover an inferior conjunction of planet d. \emph{TESS} observed the system again during cycle 3 in sector 38 of its extended mission, from 28 April to 26 May 2021. Both 2-minute and 20-second cadence observations are available for this new sector, which covers two transits of planet b and one transit of planet c. However, these observations again did not cover a transit of planet d, according to the ephemeris of \citet{delrez2021}. The data were processed by the Science Processing Operations Center (SPOC) pipeline at NASA Ames Research Center \citep{jenkins2010,jenkins2016}. We retrieved the presearch data conditioning simple aperture photometry (PDC\_SAP, \citealt{stumpe2012,smith2012,stumpe2014}) from the Mikulski Archive for Space Telescopes (MAST\footnote{\url{https://archive.stsci.edu/}}) and removed all data points for which the quality flag was not zero. 

\begin{figure*}[!h]
    \centering
    \resizebox{\textwidth}{!}{\includegraphics{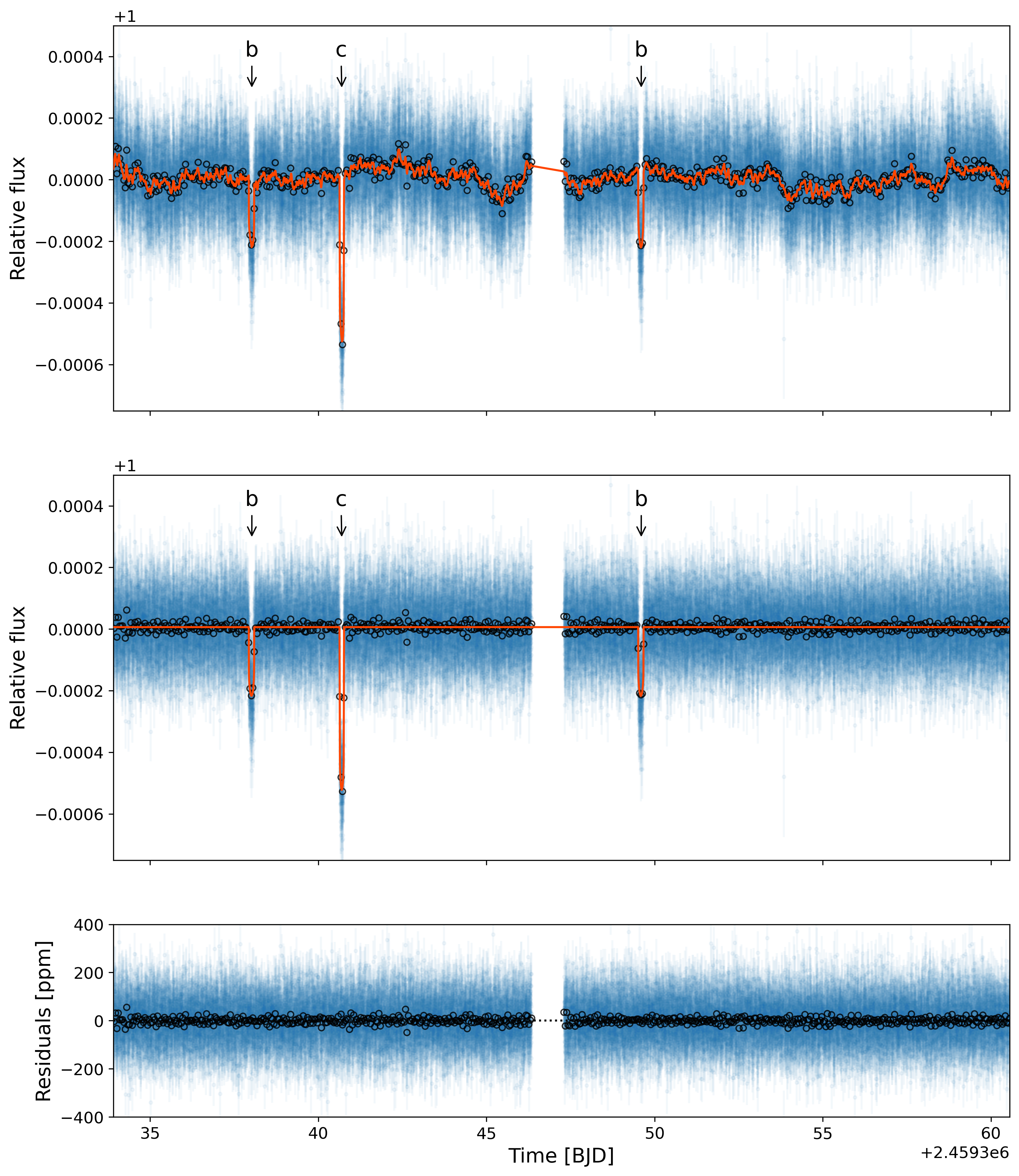}}
    \caption{\emph{TESS} sector 38 light curve of \nulup\ . \emph{Top panel:} Raw light curve obtained by binning the 20-second cadence data into 2-minute intervals (blue points with error bars), together with our best-fit model (orange curve), which includes the transits of planets b and c, and the GP model used to account for the correlated noise. \emph{Middle panel:} Light curve obtained after subtracting the GP component of our model, together with our best-fit transit model (orange curve). \emph{Bottom panel:} Residuals resulting from the difference between the detrended light curve and the transit model. For all panels, the error bars of the data points include the fitted additional jitter term added in quadrature. The open black circles show the light curve binned into one-hour intervals.}
    \label{fig:TESS}
\end{figure*}

\cite{huber2022} recently reported that \emph{TESS} 20-second light curves show a $\sim$10-25\% improvement in photometric precision compared to the 2-minute data for bright stars ($T$-mag $\lesssim$ 8), when binned to the same cadence. This is consistent with pre-flight expectations and related to differences in the cosmic-ray rejection algorithms applied to the 20-second and 2-minute data. According to their Fig.~2, this improvement can even be as large as $\sim$30\% for very bright stars with $T$-mag$\sim$5, such as \nulup\ ($T$-mag=5.05). Checking this for the \nulup\ PDC\_SAP photometry, we indeed found a $\sim$29\% improvement when comparing the RMS of the 20-second light curve binned into 2-minute intervals (103 ppm) with the one of the 2-minute light curve (146 ppm). We thus decided to use the binned 20-second light curve in our analysis. This light curve consists of 18\,482 data points and is shown in the top panel of Fig.~\ref{fig:TESS}.

\subsection{Archival data}
To derive the strongest constraints on the system parameters, we also included the data previously published in \cite{udry2019}, \cite{kane2020}, and \cite{delrez2021} in our combined analysis. These data sets are (i) the 246 radial velocities obtained between 27 May 2004 and 4 August 2017 with the HARPS spectrograph on ESO 3.6~m telescope (La Silla, Chile), (ii) the \emph{TESS} 2-minute-cadence photometry obtained during sector 12 of its primary mission (21 May-18 June 2019), covering two transits of planet b and one transit of planet c, and (iii) the six \cheops\ visits obtained between 4 April and 6 July 2020 and covering in total four transits of planet b (one of which is partial), three transits of planet c, and one partial transit of planet d.
We refer to the studies cited above for more details about these data and their reduction. A log of the photometric observations can be found in Table~\ref{tab:obslog}.

\begin{table*}[]
\caption{\label{tab:obslog} Log of photometric observations. The spectroscopic observations with HARPS at ESO~3.6m are those from \citet{udry2019}. \emph{TESS} sector~12 data and \cheops\ 2020 data have previously been reported by \citet{kane2020} and \citet{delrez2021}, respectively.}
\begin{tabular}{lllrrrrl}
\hline
         & Data ID                              & Start date          & Duration & Frames  & $N_\mathrm{exp}\times T_\mathrm{exp}$ & Efficiency$^b$ & Planet(s) \\
         &                                      & (UTC)               & (h)      &         & (s)                                   & (\%)       & in transit          \\           
\hline
\cheops\ & CH\_PR100041\_TG000901\_V0200        & 2020-04-04T15:07    & 11.58    & 558     & $26\times1.7^a$               & 59.2            & b         \\ 
\cheops\ & CH\_PR100041\_TG000101\_V0200        & 2020-04-14T16:23    & 10.99    & 567     & $26\times1.7$               & 63.3            & c         \\ 
\cheops\ & CH\_PR100041\_TG000902\_V0200        & 2020-04-16T03:59    & 11.62    & 580     & $26\times1.7$               & 61.2            & b         \\ 
\cheops\ & CH\_PR100041\_TG001101\_V0200        & 2020-04-27T18:00    & 12.84    & 661     & $26\times1.7$               & 63.2            & b         \\ 
\cheops\ & CH\_PR100041\_TG001001\_V0200        & 2020-06-08T21:33    & 11.65    & 551     & $26\times1.7$               & 58.1            & c, d      \\ 
\cheops\ & CH\_PR100041\_TG001501\_V0200        & 2020-07-06T10:40    & 11.56    & 496     & $26\times1.7$               & 52.7            & b, c      \\    
\cheops\ & CH\_PR100031\_TG039201\_V0200        & 2021-04-24T22:11    & 92.11    & 4\,720  & $26\times1.7$               & 62.9            & d         \\ 
\emph{TESS} & S12-0000000136916387-0144-s       & 2019-05-21T11:00    & 672      & 20\,119 & $1\times120$                & 99.8            & $2\times$b, $1\times$c      \\
\emph{TESS} & S38-0000000136916387-0209-a\_fast & 2021-04-29T08:35    & 648      & 18\,482$^c$ & $1\times20$             & 95.1            & $2\times$b, $1\times$c      \\
\hline \hline
\end{tabular}
\newline
{\footnotesize 
$^a$ Each \cheops\ subarray image results from the on-board co-addition of 26 exposures of 1.7~s. \\
$^b$ The efficiency quantifies the interruption of \cheops\ observations due to Earth occultations and passages through the South Atlantic Anomaly. It is calculated as $N_\mathrm{exp}\times  T_\mathrm{exp}\times \mathrm{frames} / \mathrm{duration}$. \\
$^c$ After binning the 20~s exposures to the 2~min cadence.}
\end{table*}

\section{Analysis}

\subsection{Stellar properties}
\label{sec:star}

We determined the radius of \nulup\ (\object{\emph{Gaia} EDR3 5902750168276592256}) in a similar manner as in \citet{delrez2021}, using updated \emph{Gaia} EDR3 photometry and parallax, which allowed us to refine the radius value compared to the literature. In brief, we used a Markov Chain Monte Carlo (MCMC) modified infrared-flux method \citep[IRFM;][]{blackwell1977,schanche2020} to compute the bolometric flux of the target by fitting \emph{Gaia}, 2MASS, and \emph{WISE} broadband photometry \citep{gaia2021,skrutskie2006,wright2010} with stellar atmospheric models \citep{castelli2003} and converting the bolometric flux into stellar effective temperature and angular diameter. Using the offset-corrected \emph{Gaia} EDR3 parallax \citep{lindegren2021}, we determine the stellar radius of \nulup\ to be $1.054\pm0.014$~\rsun .

The stellar radius $R_{\star}$ together with the effective temperature $T_{\mathrm{eff}}$ and metallicity [Fe/H] \citep{sousa2008,delrez2021} constitute the basic input set to then derive the isochronal mass $M_{\star}$ and age $t_{\star}$, for which we employed two different stellar evolutionary models. A first pair of mass and age values was computed through the isochrone placement algorithm \citep{bonfanti15,bonfanti16} using its capability of fitting the input parameters to pre-computed grids of the \texttt{Padova and Trieste stellar evolutionary code} \citep[\texttt{PARSEC}\footnote{\url{http://stev.oapd.inaf.it/cgi-bin/cmd}};][]{marigo17} isochrones and tracks. The stellar $v\sin{i}$ was also added to the basic input set to improve the routine convergence as detailed in \citet{bonfanti16}. 
To compute the second pair of mass and age values, we applied another code, namely the \texttt{code li\'egeois d'\'evolution stellaire} \citep[\texttt{CLES};][]{scuflaire08}. It uses the constraints given by the input parameters to produce the best-fit evolutionary tracks of the target star following the Levenberg-Marquadt minimisation scheme \citep[see][for further details]{salmon21}.
After confirming the mutual consistency of the two respective pairs of outcomes through the $\chi^2$-based criterion described in \citet{bonfanti21}, we finally merged the pairs of output distributions and obtained $M_{\star}=0.876_{-0.032}^{+0.026}$~M$_{\odot}$ and $t_{\star}=11.7_{-2.3}^{+2.1}$ Gyr. The uncertainties on $M_\star$ and $t_\star$ were propagated from the uncertainties on $R_\star$, $T_\mathrm{eff}$ and [Fe/H], as described in \citet{bonfanti15} and \citet{bonfanti16} for the isochrone placement and in \citet{scuflaire08} and \citet{salmon21} for the CLES code. All stellar parameters we used and derived are listed in Table~\ref{tab:star}.

\subsection{Combined data analysis}
\label{sec:combinedDataAnalysis}

We performed a combined analysis of the whole data set using the \texttt{juliet} python package \citep{Espinoza2019}, which is built on several publicly available tools such as \texttt{batman} \citep{kreidberg2015} for the modelling of transits, \texttt{radvel} \citep{fulton2018} for radial velocities, \texttt{celerite} \citep{Foreman-Mackey2017} for Gaussian processes (GPs), and \texttt{dynesty} \citep{speagle2020} for estimating Bayesian posteriors and evidence via dynamic nested sampling.
Our analysis is very similar to the one presented in \cite{delrez2021} and simply incorporating the new \cheops\ and \emph{TESS} light curves. 

The fitted system parameters were, for each planet: the orbital period $P$, the mid-transit time $T_{0}$, the radial velocity semi-amplitude $K$ and the parameters $r_1$ and $r_2$, which are connected to the planet-to-star radius ratio $p$ and the transit impact parameter $b$ via equations (1)-(4) in \cite{espinoza2018}. This parametrisation was shown to allow an efficient sampling of the physically plausible values in the ($b, p$) plane. We parametrised the stellar density $\rho_{\star}$ which, together with the orbital period $P$ of each planet, defines a value for the scaled semi-major axis $a/R_{\star}$ of each planet through Kepler's third law, where $R_\star$ is the stellar radius. This parametrisation offers the advantage of defining a single common value of the stellar density for the system, rather than fitting for the scaled semi-major axis of each planet, thus reducing the number of fitted parameters. We placed a normal prior $\mathcal{N}(\mathrm{\mu=1\,052~kg~m^{-3},~\sigma^2=(56~kg~m^{-3})^2})$ (i.e. $\rho_{\star}=0.746\pm0.041\,\rho_{\odot}$) on the stellar density based on the stellar mass ($M_{\star}=0.876_{-0.032}^{+0.026}$~M$_{\odot}$) and radius ($R_{\star} = 1.054 \pm 0.014$~R$_{\odot}$) that we derived previously (Sect. \ref{sec:star}). Finally, for each band pass (\cheops\ and \emph{TESS}), two quadratic limb-darkening coefficients were parametrised using the ($q_1$, $q_2$) triangular sampling scheme of \cite{kipping2013}. All these parameters were sampled from wide uniform priors, except for the stellar density (see above). We assumed circular orbits for the three planets, as justified in \cite{delrez2021}.

We modelled the correlated noise present in the light curves simultaneously with the planetary signals to ensure a full propagation of the uncertainties. To this end, we first performed individual analyses of each light curve in order to select the best correlated noise model for each of them based on Bayesian evidence. The new \cheops\ visit shows some typical flux variations phased with the spacecraft roll angle $\phi$ \citep[also seen for all previous visits;][]{delrez2021}, which we modelled using linear functions of sin($n\phi$) and cos($n\phi$), where $n$ = 1, 2, 3 (blue model in Fig.~\ref{fig:lc_d_CHEOPS}). In addition, the \cheops\ light curve also shows some higher-frequency stellar noise, which we modelled using a GP with a stochastically driven damped simple harmonic oscillator (SHO) kernel \citep{Foreman-Mackey2017}, with a quality factor of $1/\sqrt{2}$. As described in \cite{delrez2021}, this stellar variability is seen in all the \cheops\ light curves, therefore we fitted a single common SHO GP across the seven \cheops\ visits in our combined analysis (green model in Fig.~\ref{fig:lc_d_CHEOPS}). We modelled the correlated noise in \emph{TESS} sector 38 light curve with an exponential GP \citep{Foreman-Mackey2017} and used the noise models detailed in \cite{delrez2021} for the archival data. Finally, we also fitted an additional jitter term for each of our photometric and radial velocity data sets that was added quadratically to the error bars of the data points to account for any underestimation of the uncertainties or any excess noise not captured by our modelling.

The best-fit models for the new \cheops\ and \emph{TESS} light curves are shown in Figs.~\ref{fig:lc_d_CHEOPS} and~\ref{fig:TESS}, respectively. A close-up of the \cheops\ transit of \nulup~d is shown in Fig.~\ref{fig:lc_zoom}. The medians and 1$\sigma$ credible intervals of the system parameter posterior distributions are given in Table \ref{tab:transit_parameters}. The posterior distributions of all parameters are shown as a corner plot in Fig.~\ref{fig:cornerplot}.

\begin{table}[]
\caption{Stellar parameters of \nulup .}
\label{tab:star}
    \centering
    \begin{tabular}{lc}
\hline
\multicolumn{2}{l}{Designations} \\
\hline
\multicolumn{2}{l}{\emph{Gaia} EDR3 5902750168276592256} \\
\multicolumn{2}{l}{HD~136352} \\
\multicolumn{2}{l}{TOI~2011} \\
\multicolumn{2}{l}{TIC~136916387} \\
\hline
Parameter                   & Value \\
\hline
$\alpha$                    & $\rm 15^h 21^m 48.18^s$       \\
$\delta$                    & $\rm -48^\degree 19\arcmin 03\farcs$38      \\
Distance (pc)               & 14.7 \\
Spectral type               & G4{\sc v} \\
$V$ mag                     & 5.65 \\
$G$ mag                     & 5.48 \\
$T$ mag                     & 5.05 \\
$J$ mag                     & 4.51 \\
$H$ mag                     & 4.16 \\
$K$ mag                     & 4.16 \\
$T_\mathrm{eff}$ (K)        & $5\,664\pm61$ \\
$\log [g \mathrm{(cm~s^{-2})}]$                    & $4.39\pm0.11$ \\
$\rm Fe/H$                  & $-0.34\pm0.04$ \\
$M_\star$ (\msun)           & $0.876^{+0.026}_{-0.032}$ \\
$R_\star$ (\rsun)           & $1.054\pm0.014$           \\
$\rho_\star$ ($\rho_\odot$) & $0.746\pm0.041$           \\
$v \sin i$ (km~s$^{-1}$)    & $<1$                      \\
age $t_\star$ (Gyr)         & $11.7^{+2.1}_{-2.3}$      \\
\hline\hline
\end{tabular}
\end{table}

\begin{table*}[]
    \caption{Parameters of the \nulup\ planets. These are the posterior values resulting from the combined data analysis (Sect.~\ref{sec:combinedDataAnalysis}).}
    \label{tab:transit_parameters}
    \centering
    \begin{tabular}{lcc}
    \hline
    Parameter & Delrez et al.\ (2021) & \textbf{This work}  \\
    \hline
    \multicolumn{3}{l}{\textbf{\nulup\ A}} \\
    $\rho_\star$ ($\rho_\odot$)         & $0.761\pm0.045$                  & $0.758\pm0.040$                     \\
    \hline
    \multicolumn{3}{l}{\textbf{\nulup\ b}} \\
    $R_b/R_\star$                       & $0.01442^{+0.00027}_{-0.00028}$  & $0.01430^{+0.00023}_{-0.00024}$     \\
    $R_b$ (\rearth)                     & $1.664\pm0.043$ (2.6\%)          & $1.643\pm0.035$ (2.1\%)             \\
    $b$ ($R_\star$)                     & $0.52^{+0.04}_{-0.05}$           & $0.505^{+0.028}_{-0.029}$           \\
    $P$ (d)                             & $11.57797^{+0.00008}_{-0.00013}$ & $11.577794^{+0.000023}_{-0.000025}$ \\
    $T_c - 2\,450\,000$ ($\rm BJD_{TDB}$) & $8\,944.3726^{+0.0015}_{-0.0017}$ & $8\,944.37064^{+0.00068}_{-0.00070}$  \\
    Transit timing uncertainty in June 2022 (min) & 15.1 & 2.0 \\
    $W$ (hours)                         & $3.935^{+0.093}_{-0.058}$        & $3.964^{+0.028}_{-0.030}$ \\
    $i$ (degree)                        & $88.49^{+0.17}_{-0.15}$          & $88.53\pm0.11$ \\
    $a_b$ (au)                            & $0.0964\pm0.0028$                & $0.0963\pm0.0021$ \\
    $M_b$ (\mearth)                     & $4.72\pm0.42$                    & $4.68\pm0.40$             \\
    $\rho_b$ (\rhoearth)                & $1.02^{+0.13}_{-0.12}$           & $1.06^{+0.12}_{-0.11}$      \\
    \hline
    \multicolumn{3}{l}{\textbf{\nulup\ c}} \\
    $R_c/R_\star$                       & $0.02526^{+0.00047}_{-0.00044}$    & $0.02485^{+0.00038}_{-0.00037}$    \\
    $R_c$ (\rearth)                     & $2.916^{+0.075}_{-0.073}$ (2.6\%)  & $2.857^{+0.058}_{-0.057}$ (2.0\%)  \\
    $b$ ($R_\star$)                     & $0.872\pm0.007$                    & $0.869\pm0.006$                    \\
    $P$ (d)                             & $27.59221\pm0.00011$               & $27.592076_{-0.000049}^{+0.000047}$     \\
    $T_c - 2\,450\,000$ ($\rm BJD_{TDB}$) & $8\,954.40990^{+0.00052}_{-0.00054}$ & $8\,954.40942_{-0.00049}^{+0.00050}$ \\
    Transit timing uncertainty in June 2022 (min) & 4.8 & 1.7 \\
    $W$ (hours)                         & $3.251^{+0.033}_{-0.031}$         & $3.272\pm0.027$ \\
    $i$ (degree)                        & $88.571^{+0.042}_{-0.045}$        & $88.580^{+0.032}_{-0.033}$ \\
    $a_c$ (au)                            & $0.1721\pm0.0050$                 & $0.1717\pm0.0037$ \\
    $M_c$ (\mearth)                     & $11.24^{+0.65}_{-0.63}$           & $11.22^{+0.60}_{-0.58}$    \\
    $\rho_c$ (\rhoearth)                & $0.453^{+0.045}_{-0.041}$         & $0.481^{+0.040}_{-0.037}$      \\
    \hline
    \multicolumn{3}{l}{\textbf{\nulup\ d}} \\
    $R_d/R_\star$                       & $0.02219^{+0.00067}_{-0.00057}$   & $0.02181\pm0.00022$                \\
    $R_d$ (\rearth)                     & $2.562^{+0.088}_{-0.079}$ (3.4\%) & $2.507\pm0.042$ (1.7\%)            \\
    $b$ ($R_\star$)                     & $0.41^{+0.14}_{-0.21}$            & $0.353^{+0.043}_{-0.050}$          \\
    $P$ (d)                             & $107.245\pm0.050$                 & $107.1363^{+0.0019}_{-0.0024}$      \\
    $T_c - 2\,450\,000$ ($\rm BJD_{TDB}$) & $9\,009.7759^{+0.0101}_{-0.0096}$   & $9\,331.18761^{+0.00100}_{-0.00096}$ \\
    Transit timing uncertainty in June 2022 (min) & 504.4 & 14.0 \\
    $W$ (hours)                         & $8.87^{+0.56}_{-0.63}$            & $9.062^{+0.054}_{-0.052}$ \\
    $i$ (degree)                        & $89.73^{+0.14}_{-0.09}$           & $89.766^{+0.036}_{-0.033}$ \\
    $a_d$ (au)                            & $0.425\pm0.012$                   & $0.4243\pm0.0092$ \\
    $M_d$ (\mearth)                     & $8.82^{+0.93}_{-0.92}$            & $8.66^{+0.90}_{-0.91}$     \\
    $\rho_d$ (\rhoearth)                & $0.522^{+0.078}_{-0.072}$         & $0.549^{+0.064}_{-0.062}$     \\
    \hline\hline
    \end{tabular}
\end{table*}

\section{Searching for a moon}

\subsection{Motivations and limitations}
\label{sec:motivations}
\nulup\ d has one of the longest known periods of all transiting planets. With an orbital period of 107.1 days and a low eccentricity, its orbit would be located between those of Mercury and Venus in the Solar System. Because it is much more massive than Mercury or Venus, \nulup~d might have retained one or several satellites within its larger Hill sphere. Assuming a circular orbit, the Hill sphere radius can be approximated as $r_H = a_d \sqrt[3]{M_d/(3M_\star)}$. For \nulup~d, $r_H = 0.0092$~au ($\sim1.4\times10^6$~km), almost identical to the Hill sphere radius of Earth (0.0098~au) and substantially larger than that of Mercury (0.0012~au) and Venus (0.0067~au). By comparison, the two inner planets b and c have $r_H$ values of 0.0017~au and 0.0040~au, respectively (Fig.~\ref{fig:hillspheres}).
Therefore, \nulup~d could have retained one or several moons. In the Solar System, the highest moon-to-planet size ratio is $\sim5$\% for icy giants (Triton to Neptune) and $\sim$27\% for terrestrial planets (Moon to Earth). For the 2.51~\rearth\ and 8.66~\mearth\ \nulup~d, whose size and mass are intermediate between those of the Earth (1~\rearth, 1~\mearth) and Neptune (3.9~\rearth, 17.2~\mearth), this range would translate into a moon size range of 0.13-0.68~\rearth. The larger end of this range roughly corresponds to a Mars-size body, whose transit might be detectable given the brightness of the star and the excellent photometric precision it allows us to reach. This motivates the search for one or several moons that transit together with the planet. 

Our observations lasted for over 84~h, roughly centred on the mid-transit time of \nulup~d. This is largely sufficient to cover the entire transit of the planet Hill sphere in front of the stellar disc, which takes 27.3~h.\footnote{In an amusing coincidence, the Hill sphere radius of \nulup~d is almost exactly equal to the value of the transit chord projected length of  \nulup~d. This is convenient for calculating the duration of the Hill sphere transit: it is simply $3\times 9.1$~h.} Because the Hill sphere of \nulup~d is larger than the star, however, our transit observation cannot probe the fraction of the Hill sphere that does not transit the star; this would correspond to highly inclined moon orbits with respect to the planetary orbital plan (Fig.~\ref{fig:hillspheres}). As can be seen in Fig.~\ref{fig:hillspheres}, most such orbits would have to be retrograde; in fact, prograde moons are only gravitationally stable out to $\sim$0.5~$r_H$ (for circular orbits), while only retrograde moons are stable out to $1~r_H$ \citep{domingos2006}. Most of the large satellites in the Solar System have prograde orbits, with the notable exception of the Neptune moon Triton. 

The existence of several apparent small transit-like features in the \cheops\ light curves has motivated several independent analyses of light curves obtained with different pipelines. We summarise two of these efforts below. These efforts converged to conclude that there is no evidence for an exomoon signal in these data. However, a series of transit observations is needed to explore the full extent of the orbital parameter space of a moon.

\subsection{Alternative photometric extractions}
\label{sec:alternativePhotometricExtractions}

In addition to the standard \texttt{DRP} reduction of \cheops\ data \citep[Sect.~\ref{sec:cheopsDataReduction};][]{hoyer2020}, which uses aperture photometry, we ran an alternative and independent pipeline based on point spread function (PSF) fitting, called the \texttt{PSF Imagette Photometric Extraction pipeline}  \citep[\texttt{PIPE}\footnote{\url{https://github.com/alphapsa/PIPE}}; see][]{morris2021,szabo2021,deline2022,brandeker2022}. We ran \texttt{PIPE} on the \cheops\ stacked subarray images (the same input as for the \texttt{DRP}) as well as on the imagettes, which are small stamps centred on the target, extracted on-board directly out of the unstacked subarray images. While the subarray images result from the on-board co-addition of 26 individual 1.7~s exposures (performed to save bandwidth), the imagettes are only stacked in groups of two, allowing us to retrieve a sampling close to the actual observing cadence. We produced two alternative light curves, one based on PSF-fitting of the subarray images, and another based on PSF-fitting of the imagettes. They are shown in Fig.~\ref{fig:PSF_PCA}. While these light curves are largely compatible with each other and with the \texttt{DRP} light curve, the \texttt{PIPE} photometry has a slightly better dispersion. More importantly, these different light-curve extractions allowed us to test for the robustness of shallow transit signatures such as those we search for below.

\subsection{Upper limits on the size of a moon from one transit of \nulup~d Hill sphere}

\begin{figure}[!h]
	\centering
	\resizebox{\columnwidth}{!}{\includegraphics[trim={0.6cm 0.2cm 1.2cm 1.0cm},clip]{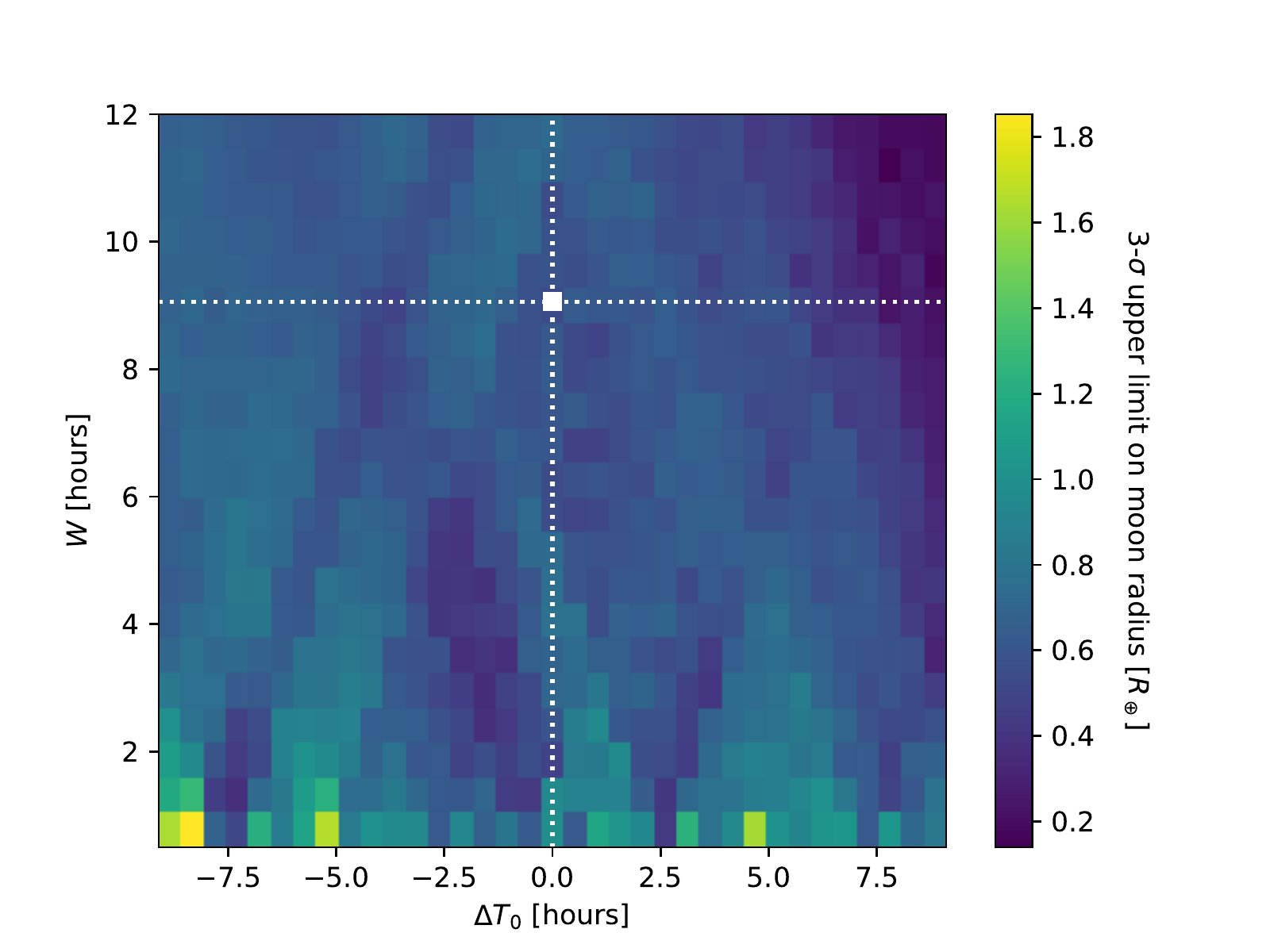}}
	\caption{3$\sigma$ upper limit on the moon radius as a function of the mid-transit time shift $\Delta T_0$ with respect to the planet and the transit duration $W$. The dotted white lines indicate the transit duration and location of planet \nulup~d. The horizontal axis spans the whole diameter of the Hill sphere (seen in projection on the stellar disc), where $\Delta T_0 < 0$ indicates that the moon lags the planet. The bottom lines of the map show short transits that could represent a moon transiting shortly before being hidden by the planet, or transiting close to the limb of the star. Such short transits allow for larger moons to exist while remaining undetectable in our single-epoch data set.}
	\label{fig:moon_size_grid}
\end{figure}

Modelling the transit light curve of an exomoon around \nulup~d while exploring the whole parameter space of its orbital properties would require the use of six (eight) parameters, assuming a circular (an eccentric) orbit: the exomoon radius, orbital period, orbital inclination, longitude of the ascending node, true anomaly at a given epoch (e.g.\ the mid-transit time of the planet), the moon-to-planet mass ratio, and for an eccentric orbit, the eccentricity and argument of periastron. The sensitivity of our moon search, based on only a single-epoch observation, is limited by the fact that the gravitational pull of the moon on the planet could induce planetary transit-timing variations and transit duration variations \citep{szabo2006,simon2007,kipping2013}, and also by the existence of possible planet-moon eclipses \citep{simon2009,kipping2011,hippke2022,gordon2022} . We thus caution that a deep, rigorous search for a moon requires additional epoch and follow-up observations \citep[see][for a \cheops -specific analysis]{simon2015}. Nevertheless, determining the time investment represented by such a future follow-up requires us to quantify the sensitivity reached in a single epoch. We therefore opted to simplify the problem by reducing the number of dimensions of our parameter space in the following way: We parametrised box-shape moon transits with the mid-transit time shift $\Delta T_0$ with respect to the planet, the transit duration $W$, and a transit depth $R_\text{moon}^2/R_\star^2$. We scanned a grid of $\Delta T_0$ and $W$ values, covering the full Hill sphere radius ($-9.1 \leq \Delta T_0 \leq +9.1$~h) and considering possible transit durations $0 < W \leq 12$~h. We measured the upper limit at $3\,\sigma$ on $R_\text{moon}$ by trying to fit additional box-shaped transits to the residuals of the \texttt{DRP} light curve and report the values in Figure~\ref{fig:moon_size_grid}.

The tighter detection limit on a moon radius is obtained for configurations with a long moon-transit duration. In these cases, it is possible to exclude moons with radii larger than $\sim0.6$~\rearth\ (roughly Mars-size) from the regions of the Hill sphere that are transiting the star (see Fig.~\ref{fig:hillspheres}). Larger moons at long periods (likely on retrograde orbits, as discussed in Sect.~\ref{sec:motivations}) could still hide in the parts of the Hill sphere that do not transit the star during our observation (they could transit at another epoch, however). Short moon transits allow for larger objects (up to $\sim 2$~\rearth) to fit into light curve gaps or that could be due to outlying points in the light curve. Short transits like this could happen for instance when a transiting moon is eclipsed by the planet, or for a moon that grazes the edge of the stellar disc.
The region of the parameter space with small $W$ and large $|\Delta T_0|$ is less likely because when the moon is far from the planet, it implies a rapidly transiting moon on the stellar disc or a moon transiting close to the stellar limb, which are not properly modelled when using box-shaped transits. A similar but independent search (see Appendix~\ref{app:seb}) yielded similar results. We also conducted further explorations of our \cheops\ light curve using more realistic transit models with simpler models for the moon orbit (see Appendices~\ref{app:akin} and~\ref{app:tom}); these efforts did not allow us to find any compelling evidence for a moon larger than Mars. A comprehensive future search will require additional epochs of observation.

\section{Conclusion}

We observed a complete transit of \nulup\ d with the \cheops\ space telescope. We substantially refined the planet transit ephemeris to $P = 107.1361^{+0.0019}_{-0.0022}$~days and $T_c = 2\,459\,009.7759^{+0.0101}_{-0.0096}$~$\rm BJD_{TBD}$, which improves by $\sim$40$\times$ on the previously reported transit-timing uncertainty (projected in June 2022). The total transiting duration is 9.1~h. The \cheops\ observations cover the transit of the Hill sphere of the planet, which is large for a transiting exoplanet, as large as that of Earth, and might host one or several transiting moons. Throughout several independent searches, we concluded that there is no evidence for additional transits of objects with radii $\ga 0.6$~\rearth ,  demonstrating that \cheops\ is precise enough to hunt for exomoons that have about the size of Mars. Nonetheless, the best candidate signals did not pass our reproducibility tests. We emphasise that even if they had, confirming the presence of an exomoon around \nulup~d would require additional epochs of observations to validate a circumplanetary orbit in any case. Such a follow up, by \cheops\ or other telescopes, is now enabled by the refined ephemeris we provide in this work. This will enable performing a deeper search of the planetary Hill sphere, or spectroscopically probing the atmosphere of this warm subneptune, which has one of the highest transit spectroscopy metrics (TSM) of its temperature class \citep[][see their Fig. 2d and discussion]{delrez2021}, including at wavelengths, such as that of the Lyman-$\alpha$ line of atomic hydrogen, that are obscured by the interstellar medium in more distant systems.


\begin{acknowledgements}
CHEOPS is a European Space Agency (ESA) mission in partnership with Switzerland with important contributions to the payload and the ground segment from Austria, Belgium, France, Germany, Hungary, Italy, Portugal, Spain, Sweden and the United Kingdom. The Swiss participation in CHEOPS has been supported by the Swiss Space Office in the framework of PRODEX and the Activités Nationales Complémentaires and the Universities of Bern and Geneva as well as the NCCR PlanetS and the Swiss National Science Foundation. The MOC activities have been supported by ESA contract 4000124370. This work has been carried out within the framework of the National Centre of Competence in Research PlanetS supported by the Swiss National Science Foundation under grants 51NF40\_182901 and 51NF40\_205606.
D.G., X.B., S.C., M.F. and J.L.\ acknowledge their roles as ESA-appointed CHEOPS science team members.
This project has received funding from the European Research Council (ERC) under the European Union’s Horizon 2020 research and innovation programme (project {\sc Four Aces}; grant agreement No 724427).  It has also been carried out in the frame of the National Centre for Competence in Research PlanetS supported by the Swiss National Science Foundation (SNSF). DE acknowledges financial support from the Swiss National Science Foundation for project 200021\_200726.
The Belgian participation to CHEOPS has been supported by the Belgian Federal Science Policy Office (BELSPO) in the framework of the PRODEX Program, and by the University of Liège through an ARC grant for Concerted Research Actions financed by the Wallonia-Brussels Federation; L.D. is an F.R.S.-FNRS Postdoctoral Researcher.
ACC and TW acknowledge support from STFC consolidated grant numbers ST/R000824/1 and ST/V000861/1, and UKSA grant number ST/R003203/1.
SH gratefully acknowledges CNES funding through the grant 837319.
This work was also partially supported by a grant from the Simons Foundation (PI Queloz, grant number 327127).
YA and MJH acknowledge the support of the Swiss National Fund under grant 200020\_172746.
ML and BA acknowledges support of the Swiss National Science Foundation under grant number PCEFP2\_194576.
S.G.S. acknowledge support from FCT through FCT contract nr. CEECIND/00826/2018 and POPH/FSE (EC).
We acknowledge support from the Spanish Ministry of Science and Innovation and the European Regional Development Fund through grants ESP2016-80435-C2-1-R, ESP2016-80435-C2-2-R, PGC2018-098153-B-C33, PGC2018-098153-B-C31, ESP2017-87676-C5-1-R, MDM-2017-0737 Unidad de Excelencia Maria de Maeztu-Centro de Astrobiolog\'\i a (INTA-CSIC), as well as the support of the Generalitat de Catalunya/CERCA programme. The MOC activities have been supported by the ESA contract No. 4000124370.
S.C.C.B. acknowledges support from FCT through FCT contracts nr. IF/01312/2014/CP1215/CT0004.
ABr was supported by the SNSA.
This project was supported by the CNES.
This work was supported by FCT - Funda\c{c}\~ao para a Ci\^encia e a Tecnologia through national funds and by FEDER through COMPETE2020 - Programa Operacional Competitividade e Internacionaliza\c{c}~ao by these grants: UID/FIS/04434/2019, UIDB/04434/2020, UIDP/04434/2020, PTDC/FIS-AST/32113/2017 \& POCI-01-0145-FEDER- 032113, PTDC/FIS-AST/28953/2017 \& POCI-01-0145-FEDER-028953, PTDC/FIS-AST/28987/2017 \& POCI-01-0145-FEDER-028987, O.D.S.D. is supported in the form of work contract (DL 57/2016/CP1364/CT0004) funded by national funds through FCT.
B.-O.D. acknowledges support from the Swiss National Science Foundation (PP00P2-190080).
MF and CMP gratefully acknowledge the support of the Swedish National Space Agency (DNR 65/19, 174/18).
DG gratefully acknowledges financial support from the CRT foundation under Grant No. 2018.2323 ``Gaseousor rocky? Unveiling the nature of small worlds''.
M.G. is an F.R.S.-FNRS Senior Research Associate.
This work was granted access to the HPC resources of MesoPSL financed by the R\'egion \^Ile-de-France and the project Equip@Meso (reference ANR-10-EQPX-29-01) of the programme Investissements d'Avenir supervised by the Agence Nationale pour la Recherche.
PM acknowledges support from STFC research grant number ST/M001040/1.
LBo, GBr, VNa, IPa, GPi, RRa, GSc, VSi, and TZi acknowledge support from CHEOPS ASI-INAF agreement n. 2019-29-HH.0.
KGI is the ESA CHEOPS Project Scientist and is responsible for the ESA CHEOPS Guest Observers Programme. She does not participate in, or contribute to, the definition of the Guaranteed Time Programme of the CHEOPS mission through which observations described in this paper have been taken, nor to any aspect of target selection for the programme.
IRI acknowledges support from the Spanish Ministry of Science and Innovation and the European Regional Development Fund through grant PGC2018-098153-B- C33, as well as the support of the Generalitat de Catalunya/CERCA programme.
GyMSz acknowledges the support of the Hungarian National Research, Development and Innovation Office (NKFIH) grant K-125015, a a PRODEX Experiment Agreement No. 4000137122, the Lend\"ulet LP2018-7/2021 grant of the Hungarian Academy of Science and the support of the city of Szombathely.
V.V.G. is an F.R.S-FNRS Research Associate.
NAW acknowledges UKSA grant ST/R004838/1.
This paper includes data collected by the \emph{TESS} mission, which are publicly available from the Mikulski Archive for Space Telescopes (MAST). Funding for the TESS mission is provided by the NASA’s Science Mission Directorate. We acknowledge the use of public TESS data from pipelines at the TESS Science Office and at the TESS Science Processing Operations Center. Resources supporting this work were provided by the NASA High-End Computing (HEC) Program through the NASA Advanced Supercomputing (NAS) Division at Ames Research Center for the production of the SPOC data products.
\end{acknowledgements}

%
%



\bibliographystyle{aa}
\bibliography{biblio}


\appendix

\section{Corner plot of the combined data analysis}

\begin{figure*}
	\centering
	\resizebox{\textwidth}{!}{\includegraphics{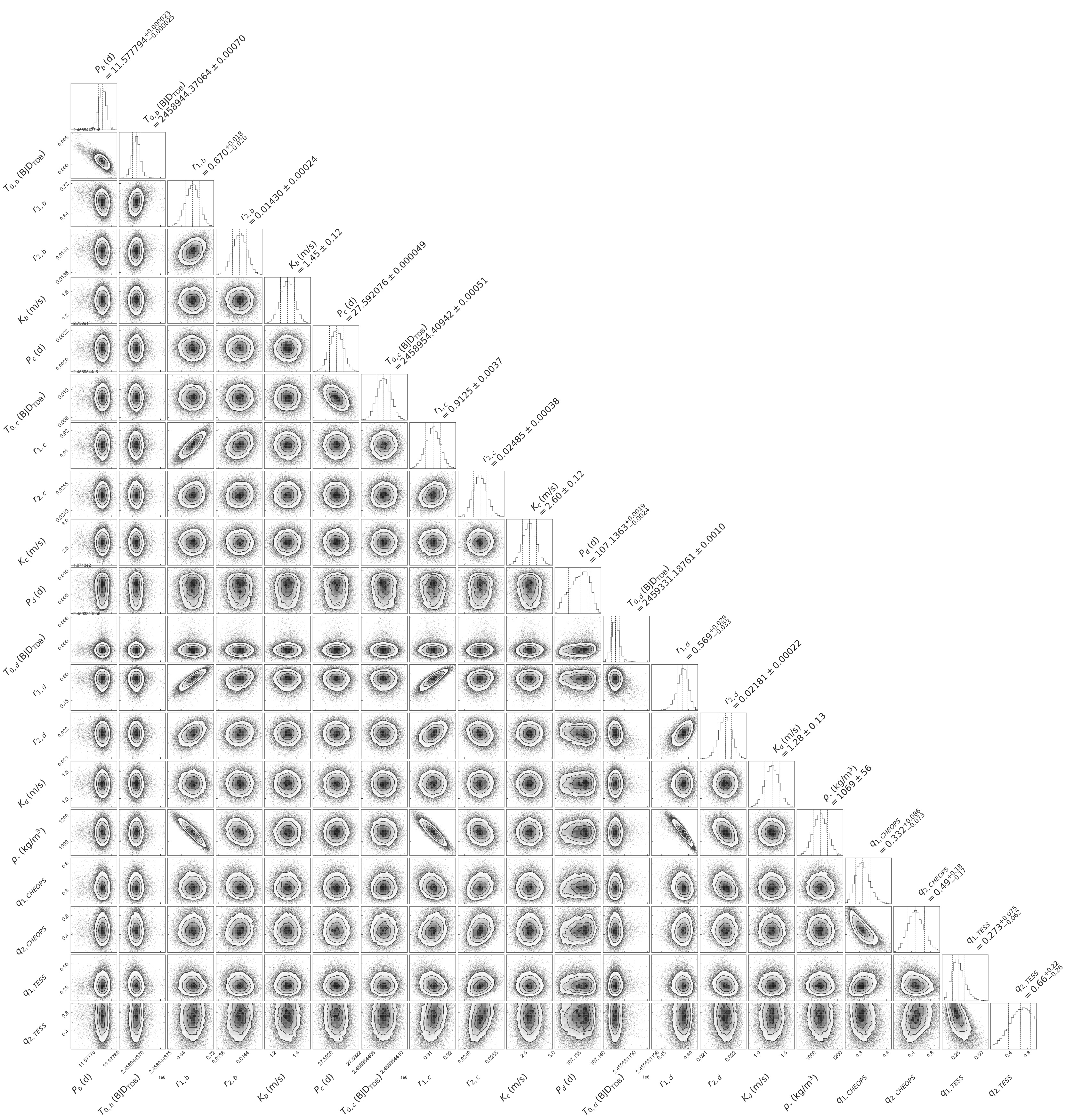}}
	\caption{Corner plot showing the posterior distributions of all parameters of the combined fit.}
	\label{fig:cornerplot}
\end{figure*}

The posterior distributions of all parameters from the combined data analysis (Sect.~\ref{sec:combinedDataAnalysis}) are shown in Fig.~\ref{fig:cornerplot}.

\section{Alternative search for moon(s) 1}
\label{app:seb}

We fitted the \cheops\ light curve with a model including a planet and a moon that took Keplerian motion and moon-planet eclipses into account. The moon was assumed (1) to have a circular orbit beyond the planet's Roche limit, but inside the planet's Hill sphere; (2) to be co-aligned with the orbit of the planet; and (3) to have an arbitrary transit depth from 10 to 1\,000~pm. The moon orbit is thus parametrised with only two parameters $(P,\theta)$: the moon orbital period $P$ and its orbital phase $\theta$ at mid-transit. This choice is motivated by the fact that our single 9.1~h planetary transit represents only a small fraction of the hypothetical moon orbit: at the Roche limit, we have $P=0.3$~d, while at the Hill sphere radius, $P=115$~d), preventing any accurate fit of the six orbital elements of the moon. We considered 100 values for $\theta$ from 0 to 360~deg. For $P$, we considered 100 values from $-115$ to $+115$~d (negative values represent a retrograde orbit). For every pair of values $(P_i,\theta_i)$, we calculated the moon transit depth $D_i$ that best fit the data following a $\chi^2$ minimisation and the value of $\chi^2_i$. 

Among the 10\,000 combinations tested, the best fit was obtained for $P=+1.766$~d, $\theta=198.36$~deg and $D=23$~ppm  ($\chi^2_{moon}=1267.79$). However, this solution does not appear better than a simple $\chi^2$ fit to the data without a moon (the null hypothesis). This null hypothesis results in $\chi^2_{no-moon}=1268.816$. The difference between the two best fits is barely visible by eye. While the `moon model' has a slightly lower $\chi^2$ than the `no-moon model', this improvement is not significant because the `moon model' has a larger number of free parameters than the `no-moon model'. Two models with different degrees of freedom can be compared with the Akaike information criterion \citep[AIC;][]{Burnham},
\begin{equation}
    \mathrm{AIC} = \chi^2+2K+\frac{2K(K+1)}{N_p-K-1},
\end{equation}
where $K$ is the number of model parameters and $N_P$ is the number of data points in the fit. The AIC can be qualitatively interpreted as the $\chi^2$, but with a penalty proportional to the number of degrees of freedom. The model with the smallest AIC is the preferred hypothesis. In this case, we obtain $\rm AIC_{moon}=1278.07$ and $\rm AIC_{no-moon}=1268.81$. The `no-moon model' is thus preferred to the `moon model'. No moon seems necessary to fit the data. We emphasise, however, that this does imply there is no moon around \nulup~d, given the limitations expressed in Sect.~\ref{sec:motivations}.

\section{Alternative search for moon(s) 2}
\label{app:akin}
We searched the light curve for additional dips that might be caused by a transiting moon. This was done by adding a second (moon) transit model to the planet transit model (both created using \texttt{batman}) to generate a simple planet-moon transit model.  The planet model parameters are those from Table~\ref{tab:transit_parameters} (right column). 

We assumed that the moon is co-moving with the planet across the stellar disc and thus has the same transit duration. Therefore, several parameters remain the same for the moon and the planet transits\citep{kipping2011} (e.g.\ we fitted for only one stellar density value). The planet and moon also share the same impact parameter that was considered to be the impact parameter of the planet-moon barycenter across the stellar surface \citep{kipping2010}. Thus, the only moon parameters we fitted for were its mid-transit time ($T_0^M$) and radius ratio ($p_M =R_M/R_\star$). We allowed for a nonphysical negative-radius moon (inverted transit) to avoid biasing towards a detection and allowed for a good sampling of values around 0. Based on the Hill radius of the planet, the maximum separation of a plausible moon from the planetary transit is $\pm0.37$~d (leading or lagging the planet). We used this range as prior bounds on the expected mid-transit time of the moon, but we also relaxed this constraint to search for any such dips outside the Hill radius that might indicate that a dip found within the Hill radius is spurious.

\begin{figure*}
	\centering
	\resizebox{\textwidth}{!}{\includegraphics{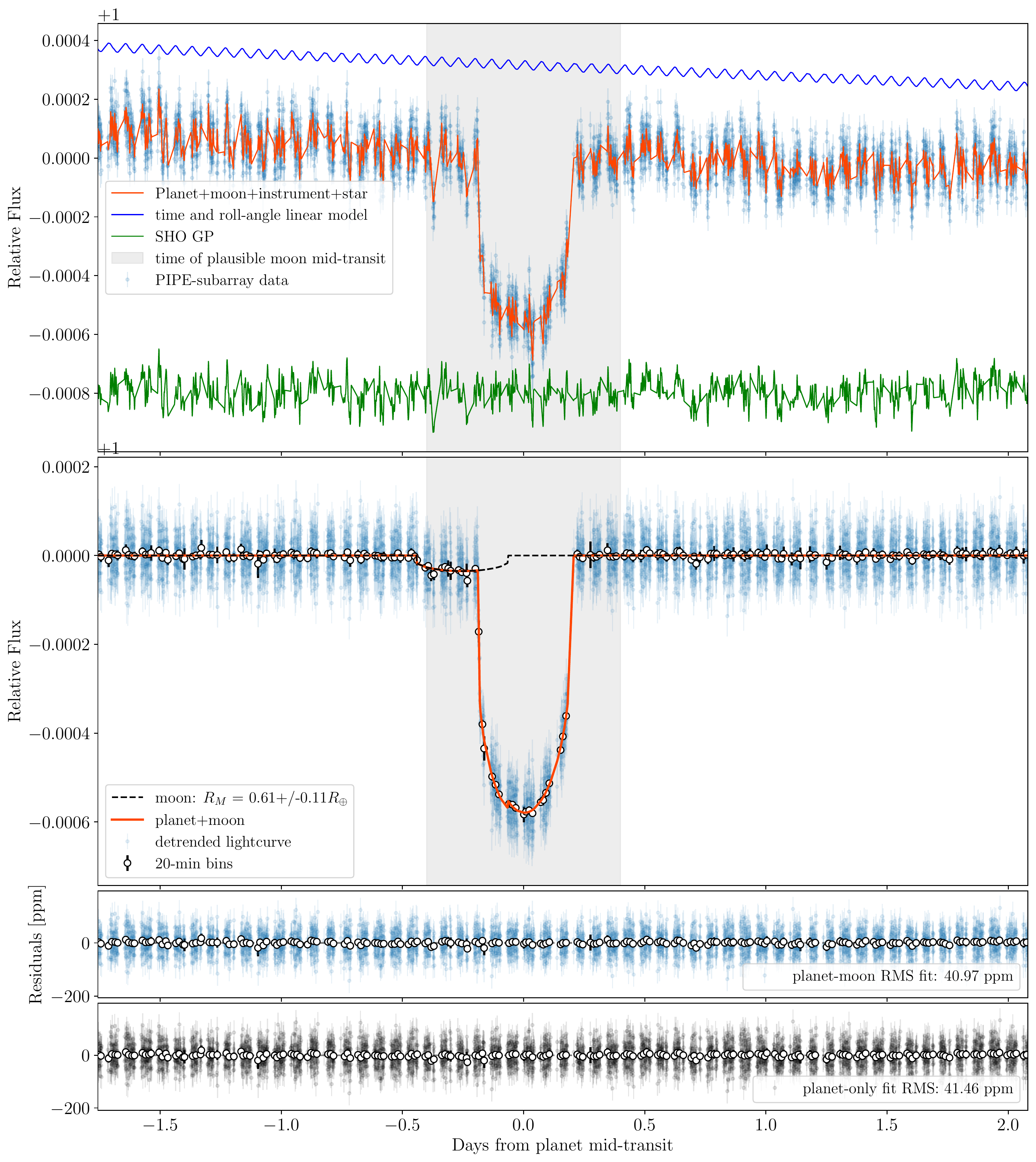}}
	\caption{Planet-moon model fit to the \emph{CHEOPS} light curve of \nulup\ . \emph{Top panel:} \texttt{PIPE}-subarray light curve (blue points with error bars) modelled with a combination of instrumental effects (blue curve), stellar noise (green curve) and transits of a planet and moon. The model including all these effects is shown in orange. The shaded grey region shows the physically plausible moon mid-transit times within the planet's Hill sphere. \emph{Middle panel:} Light curve corrected for the instrumental and stellar noise models (blue points), together with the best-fit planet-moon transit model (orange curve). The dashed black curve shows the isolated moon component. \emph{Bottom panels:} Residuals from the planet-model fit and planet-only fit. The open black circles show the light curve binned into 20-minute intervals. As described in Sect.~\ref{app:akin}, the shallow transit feature clearly appearing in the second panel from top is an artefact.}
	\label{fig:PIPE-moonfit}
\end{figure*}

We used a linear function of time and Fourier functions as in Sect.~\ref{sec:combinedDataAnalysis} to model the temporal trend and roll-angle variations in the light curve. We used the SHO kernel that was also employed to model the stellar noise (\S\ref{sec:combinedDataAnalysis}) and the \texttt{dynesty} nested-sampling routine to sample the parameter space for the planet-only and planet-moon models. This procedure was performed on the different light curve extractions mentioned above (\texttt{DRP}, \texttt{PIPE}-imagette and \texttt{PIPE}-subarray).

The analysis of the \texttt{PIPE}-subarray data reveals a dip within the Hill radius with a depth corresponding to a body of radius $0.61\pm0.1$~\rearth. This dip is found regardless of whether we constrain the search region to be within the Hill radius. Figure \ref{fig:PIPE-moonfit} shows the maximum likelihood planet-moon fit to the \texttt{PIPE}-subarray transit light curve of \nulup~d. Comparing the Bayesian evidence of the planet-moon model to the planet-only model, we obtain a Bayes factor of 9.6, which is in strong favour of the planet-moon model \citep[the threshold to reject the null hypothesis lies at a Bayes factor of $\sim$5;][]{kass1995}. In the case of the planet-only model fit, the GP is able to absorb any variation or moon-like dip that is not modelled. 

The existence of the shallow transit that is clearly visible in the GP-detrended light curve (Fig.~\ref{fig:PIPE-moonfit}, second panel) seems statistically sound based on Bayesian evidence. However, because fools rush in where angels fear to tread, we wished to further assess how reproducible and robust this signature, a product of a sophisticated detrending, really is.  \citet{rodenbeck2018} have previously shown that light-curve detrending processes can result in the injection of spurious moon transits to the data. Thus, to test whether this dip could have been induced by the GP that we used to model the stellar noise, we performed the same fit without the GP. We obtained a similar result, albeit with less precision on the transit depth. When we re-fitted the data after removing the detected shallow transit signal, we found additional dips outside the Hill sphere of \nulup~d. This indicates that a shallow transit signal found within the Hill radius could well be an artefact due to systematics (stellar noise, instrumental noise, a spurious signal created by the data reduction process or a combination of all of these). A good test for the latter scenario is to perform a similar detrending and fitting procedure to the product of the alternative data reduction pipeline mentioned in Sect.~\ref{sec:alternativePhotometricExtractions}. We might expect that different kinds of systematic effects would be created or amplified differently with two different extraction methods (aperture photometry vs.\ PSF fitting). 

This proved an efficient test to refute our candidate exomoon, as we simply were unable to retrieve the same signal in these alternative light curves. Instead, we detected several transit-like dips outside the Hill sphere, at different times from the initial candidate signal, in the \texttt{DRP} light-curve, while an inverted transit was detected in both the \texttt{PIPE} imagette and \texttt{DRP} data. The prevalence of these transit-like dips at different times in independent data reductions indicates that they are likely artefacts produced by the stellar noise, instrumental noise, the data reduction procedure or a combination of the three.

\section{Alternative search for moon(s) 3}
\label{app:tom}
\begin{figure}
	\centering
	\resizebox{\columnwidth}{!}{\includegraphics{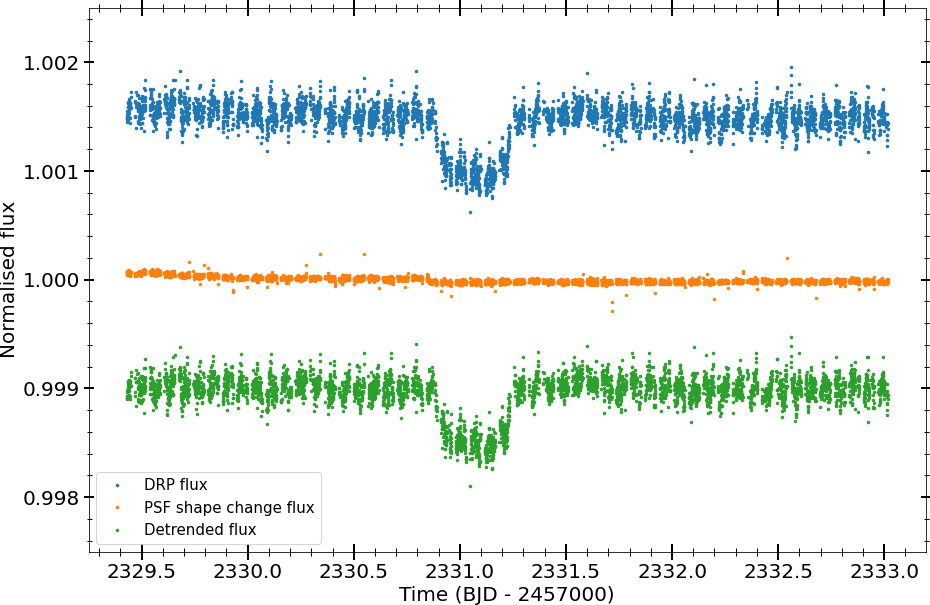}}
	\resizebox{\columnwidth}{!}{\includegraphics{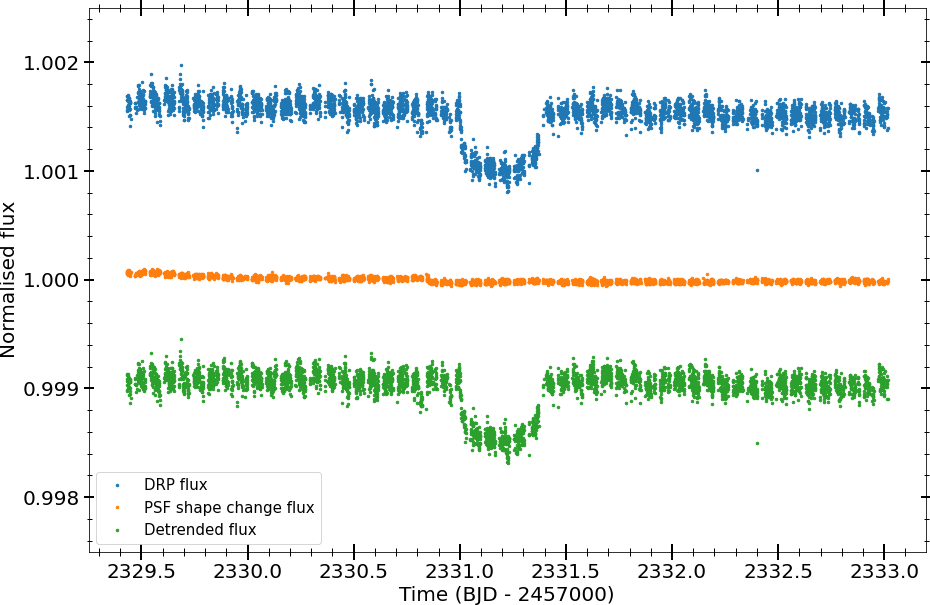}}
	\resizebox{\columnwidth}{!}{\includegraphics{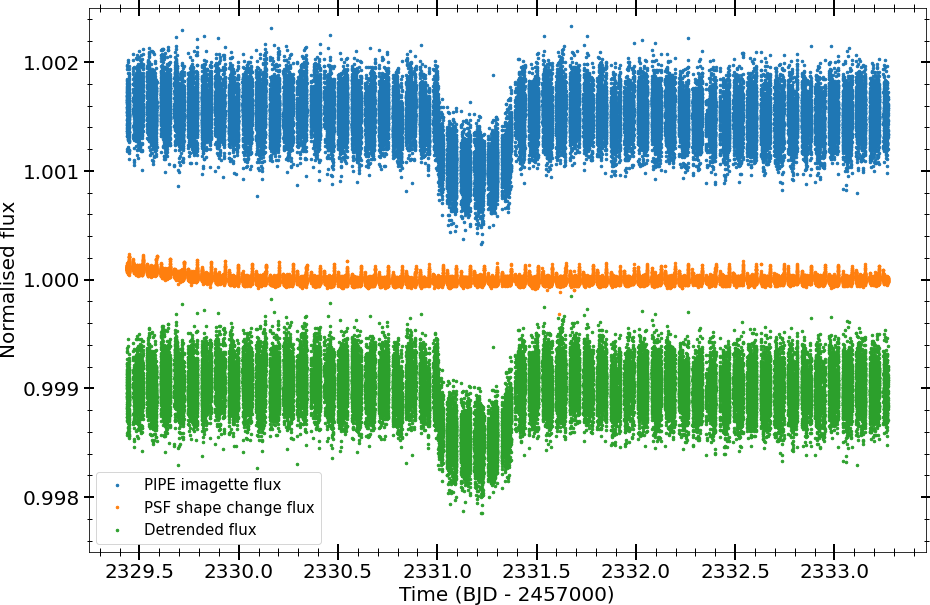}}
	\caption{Results of the PSF PCA process. \emph{Top}: Subarrays and \texttt{DRP} fluxes. \emph{Middle}: Subarrays and \texttt{PIPE} fluxes. \emph{Bottom}: Imagettes and \texttt{PIPE} fluxes. Blue represents the raw \texttt{DRP} or \texttt{PIPE} fluxes; orange shows the combined PSF shape-change noise model from the 127, 127, and 210 principal components, respectively; and green shows detrended fluxes. In the TIP analysis, we included all the individual components that comprise the orange curves and the blue raw fluxes.}
	\label{fig:PSF_PCA}
\end{figure}

A second, independent search for a shallow signal resulting from the transit of a moon similarly failed to provide convincing evidence. For this second attempt, we conducted a statistically robust analysis of the \cheops\ visit by constructing the most complete noise model to remove systematics from the data and thus optimise our search for transit-like features. The noise model contained the instrumental vectors of the visit (background, contamination, smearing, roll angle, change in temperature, and $x$ and $y$ centroid offset positions) that were retrieved using the \texttt{pycheops} Python package \citep{Maxted2021}\footnote{\url{https://github.com/pmaxted/pycheops}}. Previous \cheops\ studies have found that the telescope temperature can alter the shape of the \cheops\ PSF, which in turn produces flux variation at the beginning of a visit (the so-called `ramp'); another source of flux variations on the orbital timescale of \cheops\ is the presence of nearby contaminants \citep{Maxted2021,morris2021,Wilson2022}. To assess and remove this potential flux modulation, we used a novel PSF-based detrending method that was recently reported in \cite{Wilson2022} to remove these effects in the \cheops\ light curves of \object{TOI-1064}. We refer to that paper for the full mathematical description of the algorithm. In brief, the tool conducts a principal component analysis (PCA) on the auto-correlation function of either the \cheops\ subarrays or imagettes, depending on the light curve of interest, to assess any subtle change in the PSF shape. To find the most significant components that should be included in the overall noise model, the algorithm uses a leave-one-out cross-validation method. Fig.~\ref{fig:PSF_PCA} shows the results of this process on the \cheops\ subarrays or imagettes for the \texttt{DRP} and \texttt{PIPE} light curves. We find that the PSF PCA method \citep{Wilson2022} removes a subtle flux ramp at the beginning of the dataset.

\begin{figure}
	\centering
	\resizebox{\columnwidth}{!}{\includegraphics{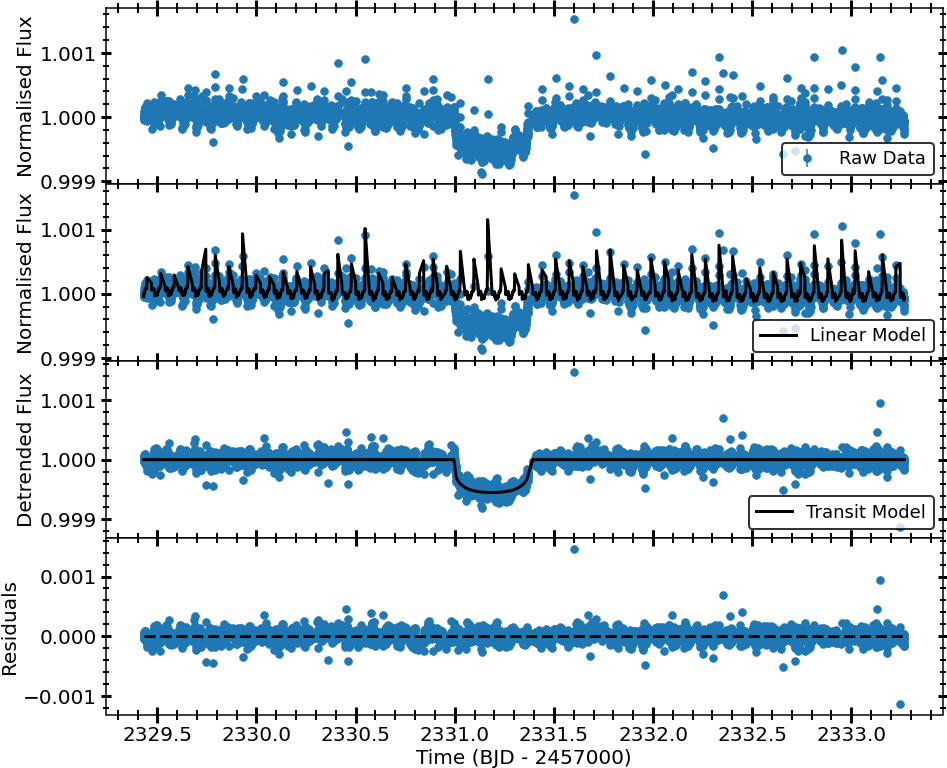}}
	\caption{Example result of the fitting method described in Sect.~\ref{app:tom} with one transit ($N=1$). \emph{Top}: \texttt{DRP} fluxes for the visit of \nulup\ d. \emph{Middle upper}: Linear model of the instrumental vectors and the components of the PSF PCA, \emph{Middle lower}: $N=1$ transit model.  \emph{Bottom}: Residuals to linear model plus $N=1$ transit model.}
	\label{fig:DRP_P=1_LC}
\end{figure}

Satisfied that we constructed a noise model that can account for as much of the systematic flux variation as possible, we fit the data with this model simultaneously with either $N=0$, 1, and 2 transit models in order to determine the true inclusion probabilities (TIP; \citealt{Hara2021}). Therefore, we can statistically verify the presence of one or two transits in the \cheops\ light curves. For the first transit model, we took priors on \nulup\ d from the results of our global analysis (see Table.~\ref{tab:transit_parameters}), whereas for the second transit model, we assumed a similar orbital period and impact parameter as \nulup\ d, and left uniform priors on the transit depth and centre time. An example result from the $N=1$ fit using the \texttt{DRP} fluxes is shown in Fig.~\ref{fig:DRP_P=1_LC}. By comparing the Bayes evidence and posterior distributions of the $N=0$ and $N=1$ fits, we find TIPs $\sim1$ for a transit at BJD $2\,459\,331.1875$ in the \texttt{DRP} light curve, and at BJD $2\,459\,331.1877$ and BJD $2\,459\,331.1876$ in the \texttt{PIPE} subarray and imagette light curves. This corresponds to the transit of \nulup~d. Confident that our process can statistically detect transits in the detrended light curves, we computed the TIP for the $N=1$ versus $N=2$ case in order to search for an additional transit that could be caused by an exomoon. For all light curves, we find TIPs $\sim0$ for all transit centre times within the data. Based on this analysis, we therefore cannot statistically confirm the presence of a transiting exomoon around \nulup\ d in these data.

\end{document}